\documentclass[10pt]{article}
\pdfoutput=1
\usepackage{jcappub}

\usepackage{url} %
\usepackage{physics} 

\usepackage{multirow}
\usepackage{booktabs}
\usepackage{placeins}

\newcommand{\unit}[1]{\,\mathrm{#1}}

\renewcommand{\vec}[1]{\boldsymbol{#1}}

\def\epsilon{\varepsilon}
\def\theta{\vartheta}
\def\lsim{\raise0.3ex\hbox{$\;<$\kern-0.75em\raise-1.1ex\hbox{$\sim\;$}}}
\def\gsim{\raise0.3ex\hbox{$\;>$\kern-0.75em\raise-1.1ex\hbox{$\sim\;$}}}

\newcommand{\apj}{{Astrophys.\ J. }}

\newcommand{\apss}{{Astrophys.\ Space Sci. }}

\newcommand{\scale}[0]{1}

\title{Revisiting cosmic ray antinuclei fluxes with a new coalescence model}

\author[a]{M. Kachelrie\ss,}
\author[b, c]{S. Ostapchenko,}
\author[a]{J. Tjemsland}

\affiliation[a]{Institutt for fysikk, NTNU, Trondheim, Norway}
\affiliation[b]{Frankfurt Institute for Advanced Studies, Frankfurt, Germany}
\affiliation[c]{D. V. Skoveltsyn Institute of Nuclear Physics, Moscow State University, Moscow, Russia}


\abstract{
  Antideuteron and antihelium nuclei have been proposed as promising detection
  channels for dark matter because of the low astrophysical backgrounds
  expected. To estimate both potential exotic contributions and their
  backgrounds, one usually employs the coalescence model in momentum
  space. Here we use instead a newly developed coalescence model based on
  the Wigner function representations of the produced nuclei states. This
  approach includes both the process-dependent size of the formation region of
  antinuclei, and the momentum correlations of coalescing antinucleons in a
  semi-classical picture. The model contains a single universal parameter
  $\sigma$ that we tune to experimental data on antideuteron production in
  electron-positron, proton-proton and proton-nucleus collisions. The obtained
  value $\sigma\simeq 1\unit{fm}$ agrees well with its physical interpretation
  as the size of the formation region of antinuclei in collisions of
  point-like particles. This model allows us therefore to calculate in a
  consistent frame-work the antideuteron and antihelium fluxes both from
  secondary production and from dark matter annihilations. We find that the
  antihelium-3 flux falls short by more than an order of magnitude
  of the detection sensitivity of the AMS-02 experiment, assuming standard
  cosmic ray propagation parameters, while the antideuteron flux can be
  comparable to the sensitivities of the AMS-02 and GAPS experiments.
}

\begin{document}

\maketitle

\section{Introduction} \label{sec:intro}

The low astrophysical backgrounds promote antideuteron~\cite{Donato:1999gy}
and antihelium-3~\cite{Cirelli:2014qia} nuclei to promising detection
channels for dark matter (DM) annihilations and decays in the Galaxy, for
a recent review see Ref.~\cite{vonDoetinchem:2020vbj}. The dominant background
of  light antinuclei is expected to originate from secondary
production, that is, to be created in collisions of primary cosmic rays (CR)
with the interstellar medium.
The high threshold energy for the production of antideuterons
($\simeq 17m_N$ in $pp$ interactions, where $m_N$ is the nucleon mass)
and antihelium-3 ($\simeq 31m_N$) implies that such secondary antinuclei have
relatively high kinetic energies. This makes antideuterons and antihelium-3
with low kinetic energies an ideal dark
matter probe. In contrast, the fluxes of heavier nuclei,
as e.g.\ antihelium-4, are, both for the DM and secondary production channels,
so strongly suppressed that they are undetectable  by current experiments.
Consequently, an identification of antihelium-4 nuclei
in the Galactic CR flux would represent a true challenge to our current
cosmological paradigm, requiring e.g.\ the presence of antimatter ``islands''
inside the Milky Way~\cite{Chardonnet:1997dv,Poulin:2018wzu}.

The production of light antinuclei as CR secondaries and in DM annihilations is
usually described by the coalescence model in momentum space~\cite{Schwarzschild:1963zz,butler_deuterons_1963,Chardonnet:1997dv}. It states that an
antiproton-antineutron pair with an invariant momentum difference $\Delta k$
less than the
coalescence momentum $p_0$ merges and forms an antideuteron. Due to the
lack of an underlying microphysical picture, $p_0$ must be determined by fits
to experimental data. For the model to be predictive, this parameter should be
independent of both the reaction type and the center-of-mass (cm) energy $\sqrt{s}$.
Traditionally, the cluster formation of nuclei has been parametrised by an
invariant coalescence factor $B_A$ as
\begin{equation}
    E_A\dv[3]{N_A}{P_A} = \left.B_A\left(E_p\dv[3]{N_p}{P_p}\right)^Z\left(E_n\dv[3]{N_n}{P_n}\right)^N\right|_{P_p=P_n=P_A/A},
\end{equation}
which relates the invariant differential yield of a nucleus with mass number
$A$, proton number $Z$ and neutron number $N$ to the invariant yields of
protons and neutrons, $E_i\dv*[3]{N_i}{P_i}$. In the limit of isotropic nucleon
yields, the coalescence factor $B_A$ is related to the coalescence momentum
$p_0$ as
\begin{equation}
    B_A=A\left(\frac{4\pi}{3}\frac{p_0^3}{m_N}\right)^{A-1}.
\end{equation}
This ``naive'' coalescence model can be
improved by taking into account two-particle momentum correlations provided by
Monte Carlo event generators, if one imposes the coalescence condition on
an event-by-event basis, as first proposed in
Refs.~\cite{kadastik_enhanced_2010,Dal_thesis}.
The yield of antinuclei should, however, depend on the full
{\em phase space density\/} of the coalescing antinucleons. Since
both  the ``naive'' and the ``improved'' coalescence models impose the
coalescence condition only in momentum space, the reaction-dependent size of
the formation region of antinuclei is neglected in these treatments. As a result,
the  coalescence parameter $p_0$ becomes process dependent
applying such models also to hadronic
reactions~\cite{aramaki_review_2016,ibarra_prospects_2013,dal_alternative_2015}. Using in such an approach the same $p_0$ for antinuclei
formation in DM annihilations and in CR interactions will thus
lead to incorrect results.

An alternative coalescence model was developed by us in
Ref.~\cite{Kachelriess:2019taq}. Starting from the Wigner function
representation of the antinucleon and the antinuclei states, 
introduced in Ref.~\cite{scheibl_coalescence_1999}, 
we employed a semi-classical treatment to include both the process-dependent 
size of the formation region and the momentum correlations of coalescing antinucleons. We showed that this new coalescence model successfully
describes the data both from  $e^+e^-$ annihilations at the $Z$
resonance~\cite{collaboration_deuteron_2006,opal_collaboration_search_1995}
and from $pp$ collisions at $\sqrt{s}=0.9, 2.7$ and
7\,TeV, measured by the ALICE collaboration
at the LHC~\cite{collaboration_production_2018}.  As we aim in the
present work to model the formation of light antinuclei as secondaries in CR
interactions, it is, however, important
to test the validity of our model also in  hadron-nucleus and light
nucleus-nucleus collisions. We consider therefore in addition  
experimental data on proton-beryllium and proton-aluminium collisions 
at $200\unit{GeV}/c$~\cite{Bussiere:1980yq,Bozzoli:1979fh}
as well as the spectra of antinuclei for $pp$ collisions at 
$\sqrt{s}=53$\,GeV measured at the 
CERN~ISR~\cite{Alper:1973my,Henning:1977mt}.
The numerical values we derive for the single free parameter $\sigma$
of our model are consistent between all the reactions considered and agree
well with the physical interpretation of $\sigma$ as the size of the formation
region of light nuclei. This allows us to calculate in a self-consistent
frame-work the expected fluxes of both antideuteron
and antihelium-3 from secondary production as well as from DM annihilations.
In the latter case, we estimate the antinuclei flux from DM particles with
masses $m_\chi = \{20, 100, 1000\}\unit{GeV}$, annihilating into $b\bar b$ and
$W^+ W^-$ pairs. We derive also the maximal annihilation cross sections
compatible with the antiproton spectrum from AMS-02.
We show that $p$He and HeHe collisions dominate the secondary contribution
to the antideuteron yield at low energies. The antihelium-3 flux we obtain
falls short of the detection sensitivities of the AMS-02 experiment by more
than an
order of magnitude, assuming standard CR propagation parameters. In contrast,
the antideuteron flux can be just below the sensitivities of the AMS-02 and GAPS
experiments. Taking into account the large uncertainties, antideuterons
remain therefore a promising target in searches for antimatter.

\section{Antinuclei formation model}
\label{sec:formation_of_antinuclei}

Our formalism for treating the production of (anti)nuclei\footnote{Since our 
discussion applies  equally well to the formation of nuclei and of antinuclei, 
we will omit the preposition `anti' further on in this section.} 
has been described  in Ref.~\cite{Kachelriess:2019taq}.
We will follow the same approach here and refer the reader for details
like cuts to exclude long-lived resonances to our previous
work~\cite{Kachelriess:2019taq}.  In this model, 
the probability that a nucleon pair with three-momentum $\vec q$ and $-\vec q$
in its cm frame coalesces is given by
\begin{equation} \label{w}
    w_\mathrm{Wigner} = 3\left(\zeta_1\Delta
    e^{-q^2d_1^2}+\zeta_2[1-\Delta]e^{-q^2d_2^2}\right),
\end{equation}
where 
\begin{equation}\label{wa}
  \zeta_i = \frac{d_i^2}{d_i^2+4\tilde\sigma^2_\perp}
            \sqrt{\frac{d_i^2}{d_i^2+4\sigma^2_\parallel}}.
\end{equation}
The parameters $\Delta=0.581$, $d_1=3.979$\,fm, and $d_2=0.890$\,fm 
determine the internal wave-function of the deuteron, which was approximated
in Ref.~\cite{Kachelriess:2019taq} as a sum of two Gaussians\footnote{The
  specified parameters correspond to the so-called $\varphi_0$-fit of the
  deuteron wave-function~\cite{Kachelriess:2019taq}.}.
Since the coalescence probability is very small, corrections to Eq.~(\ref{w}),
accounting for double counting of nucleons involved in different pairs, can
in practice be neglected. An  expression similar to  Eq.~(\ref{w}) has been
obtained in Ref.~\cite{Kachelriess:2019taq} for the probability of three
nucleons to form a bound-state, like tritium or antihelium-3.

The parameters $\sigma_i$ describe the spatial separation of the nucleons
forming potentially a deuteron. For a ``point-like'' interaction, such as
$e^+e^-$ annihilations, the longitudinal spread $\sigma_\|$ is
given in the deuteron frame by the formation length of nucleons,
$\sigma_\|\simeq R_p \simeq 1$\,fm, with $R_p$ as the proton size, while the
perpendicular spread is of order $\sigma_\perp\simeq 1/\Lambda_{\rm QCD}$ 
in the cm frame of the collision. Taking into account for the latter
the boost into the deuteron frame gives
\begin{equation} \label{LT}
  \tilde \sigma_\perp^2 =
  \frac{\sigma_\perp^2}{\cos^2\theta + \gamma^2\sin^2\theta},
\end{equation}
where $\gamma$ is the usual Lorentz factor, while $\theta$ denotes the
angle between the antideuteron momentum and the momentum of the
initially produced pair of particles in their cm frame. For instance,
in the case of the annihilation of DM particles through the process
$\chi\chi\to \bar bb$, the angle $\theta$
is defined with respect to the momentum of the produced  $b$ or $\bar b$.
For hadronic events, $\theta$ is defined relative to the beam direction
of the colliding hadrons in their cm system, as detailed in
Ref.~\cite{Kachelriess:2019taq}.

In addition, the spreads $\sigma_i$ obtain a geometrical contribution
$\sigma_\mathrm{geom}$ in reactions involving hadrons or nuclei because of
their finite extension. Adding these two contributions in quadrature yields
\begin{align}
    \sigma_\perp^2 &=\sigma_{\perp(e^{\pm})}^2 +\sigma_\mathrm{\perp(geom)}^2,\\
    \sigma_\parallel^2 &=\sigma_{\parallel(e^{\pm})}^2 +\sigma_\mathrm{\parallel(geom)}^2 . 
\end{align}
Here, we have denoted with $\sigma_{(e^{\pm})}$ the ``point-like'' contribution
discussed above and set for simplicity
$\sigma_{(e^{\pm})}\simeq \sigma_{\perp(e^{\pm})}\simeq  \sigma_{\|(e^{\pm})} \simeq 5\unit{GeV^{-1}}\simeq 1$\,fm. 
The geometrical contributions in hadron-hadron, hadron-nucleus, and nucleus-nucleus 
collisions can in turn be approximated
by~\cite{Kachelriess:2019taq}
\begin{align}
    \sigma_{\perp(\mathrm{geom})}^2 &\simeq \frac{2R_1^2R_2^2}{R_1^2+R_2^2},
    \label{eq:geom_perp}  \\
    \sigma_{\parallel(\mathrm{geom})}^2 &\simeq \mathrm{max}\{R_1^2, R_2^2\},
    \label{eq:geom_estimation}
\end{align}
where $R_i$ are the radii of the two colliding particles.
In the particular case of proton-proton collisions,
$\sigma_{\parallel}\simeq\sigma_{\perp}$ so that
$\sigma_{(pp)}\simeq\sqrt{2}\sigma_{(e^{\pm})}\simeq 7\unit{GeV^{-1}}$. 
The radius $R_A$ of a nucleus with mass number $A$ scales approximately as
\begin{equation}
    R_A = a_0 A^{1/3},
    \label{eq:scaling}
\end{equation}
where $a_0\simeq 1.1$\,fm, with an uncertainty of
$\sim 20\%$~\cite{Donnelly:2017aaa}. We will use this relation 
in Eqs.~(\ref{eq:geom_perp}) and (\ref{eq:geom_estimation}) as an
approximation for the size of the different nuclei considered. This allows us
to use a single parameter, setting
\begin{equation} \label{sigdef}
\sigma\equiv  \sigma_{(e^\pm)}  =a_0 =\sigma_{(pp)}/\sqrt{2}  .
\end{equation}
If our model accounts correctly for the differences in the formation of
light nuclei in different reaction types, the parameter $\sigma$ obtained
from fits of different reactions should be universal and close to
1\,fm.

Finally, we want to comment briefly on the relation of our model
to other approaches. The recent work~\cite{Blum:2019suo} connects
the production of light antinuclei to the two-proton correlation
function measured in heavy-ion collisions. Its basic results
can be recovered in our approach imposing two assumptions: First, 
the size of the production region has to be much larger than the
deuteron size. Second, the proton-neutron density matrix has to
factorise, i.e.\ their momentum correlations should be negligible.
Neither of the two assumptions are justified for the small systems,
as $pp$ scattering and $e^+e^-$ or DM annihilations, we consider here.
Eventually, one may ask how the size $\sigma$ of the production region
is connected to the parameter $p_0$ used in the conventional coalescence
picture. Formally, we note that $\sigma$ is approximately related to
$p_0$ as  $p_0/0.2\,\mathrm{GeV} \sim 1\,\mathrm{fm}/\sigma$. Note, however,
that important physical inputs like the shape of the momentum distributions
of antinucleons or the wave-function of the antideuteron affect $p_0$ and
$\sigma$ differently. Therefore such a relation has to be interpreted with
care.

\section{Determination of the spread $\sigma$}
\label{sec:calibration}

In order to test the validity of our coalescence model, i.e., in particular
the universality of its parameter $\sigma$, we compare our predictions 
to experimental data on antideuteron production in $e^+e^-$, $pp$
and $pA$ collisions. Differences between the results of the standard and 
the new coalescence models were already investigated  
in Ref.~\cite{Kachelriess:2019taq}, using as event
generator PYTHIA. Here we focus instead on the new model, using the
event generator QGSJET-II~\cite{Ostapchenko:2010vb,Ostapchenko:2013pia},
which reproduces experimental data over a wide energy range, for reactions
involving nuclei as well as for  $pp$ collisions\footnote{We are using
  a new tune of QGSJET-II-04m~\cite{Kachelriess:2015wpa} which slightly
  improves the fit to the ALICE data~\cite{Adam:2015qaa} on antiproton
  production in $pp$ collisions at $\sqrt{s}=7$\,TeV.}. In addition, we employ
PYTHIA~8.230~\cite{sjostrand_PYTHIA_2006,sjostrand_introduction_2015}
to simulate $e^+e^-$ and DM annihilations as well as $pp$ collisions. The
considered experimental data sets are described in
Appendix~\ref{sec:experiments}.
In Fig.~\ref{fig:pN}, we show the best fits to the data on antideuteron
production in  $p$Al and $p$Be  collisions at 200\,GeV/$c$, while the fits to
the transverse momentum $p_T$ spectra of  antideuterons in $pp$
interactions, measured at CERN~ISR and LHC, are plotted in Fig.~\ref{fig:pp}.
Because of the relatively large experimental uncertainties, the fits are in
all cases acceptable. The corresponding fit results for the parameter $\sigma$
obtained using QGSJET-II are listed in Table~\ref{tab:res_qgsjet}, while the
results for PYTHIA are shown in Table~\ref{tab:res_PYTHIA}. 
The values obtained for $\sigma$ using PYTHIA have a
smaller variance and are closer to the expected value of
$\sigma\simeq 1$\,fm, compared to the results for QGSJET-II.

\begin{figure}[htbp]
    \centering
   \includegraphics[width=0.49\columnwidth]{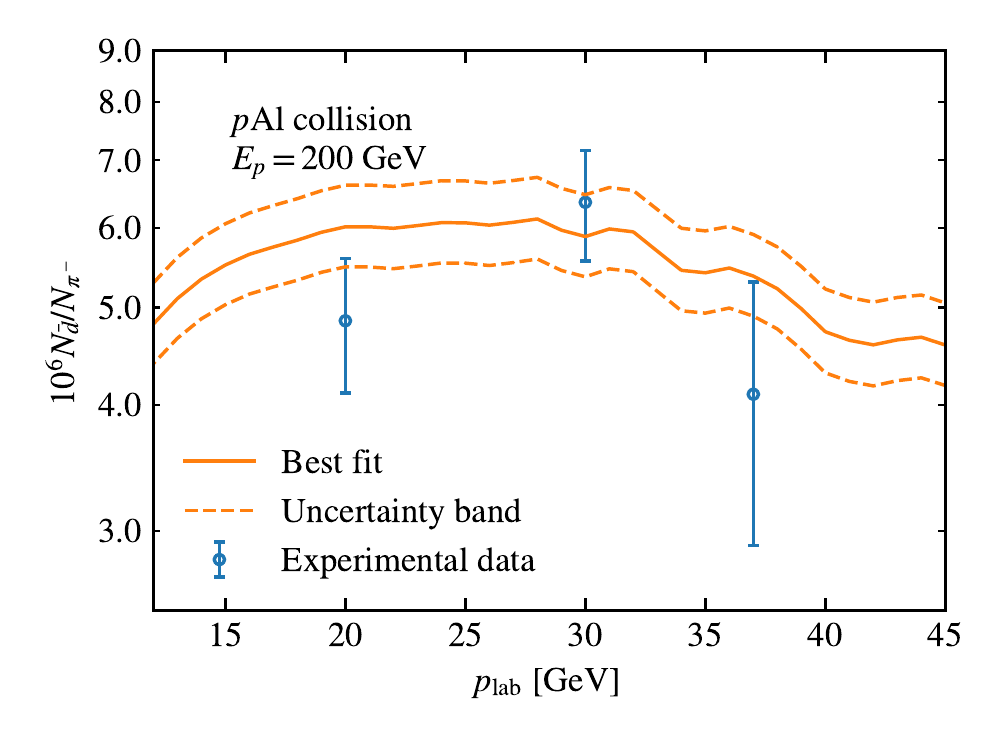}
   \includegraphics[width=0.49\columnwidth]{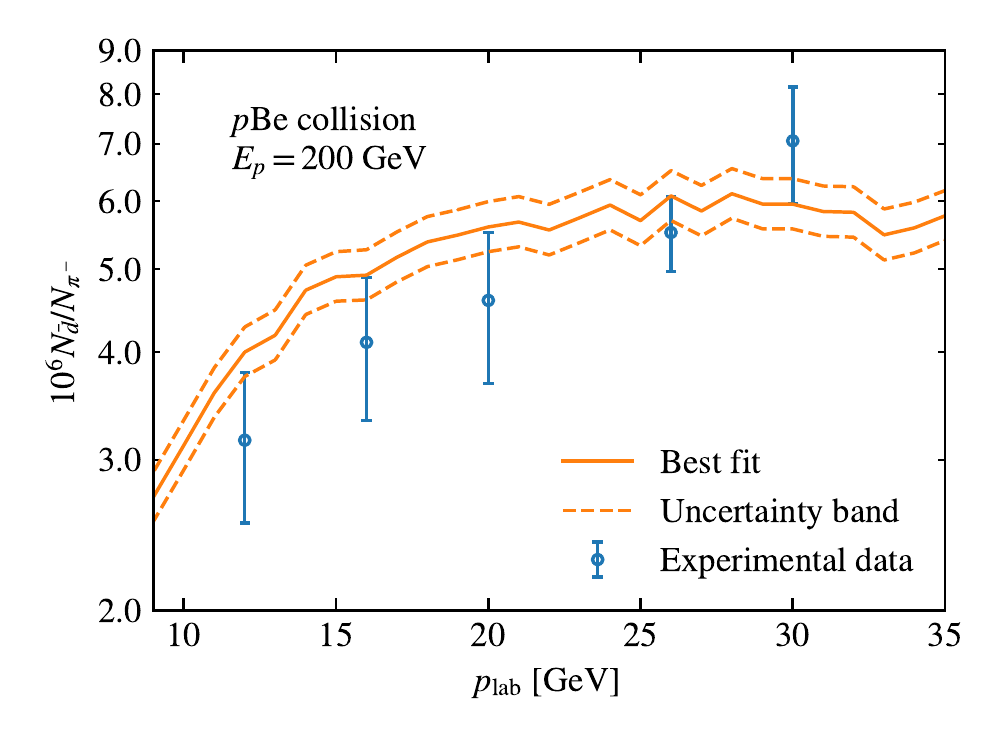}
    \vskip-0.3cm
    \caption{Best fit and $1\sigma$ uncertainty band for the antideuteron/pion
      ratio, obtained using QGSJET-IIm and the new coalescence model, for
      proton-aluminium (left) and proton-beryllium (right) collisions.} 
    \label{fig:pN}
\end{figure}

\begin{figure}[htbp]
    \centering
   \includegraphics[width=0.49\columnwidth]{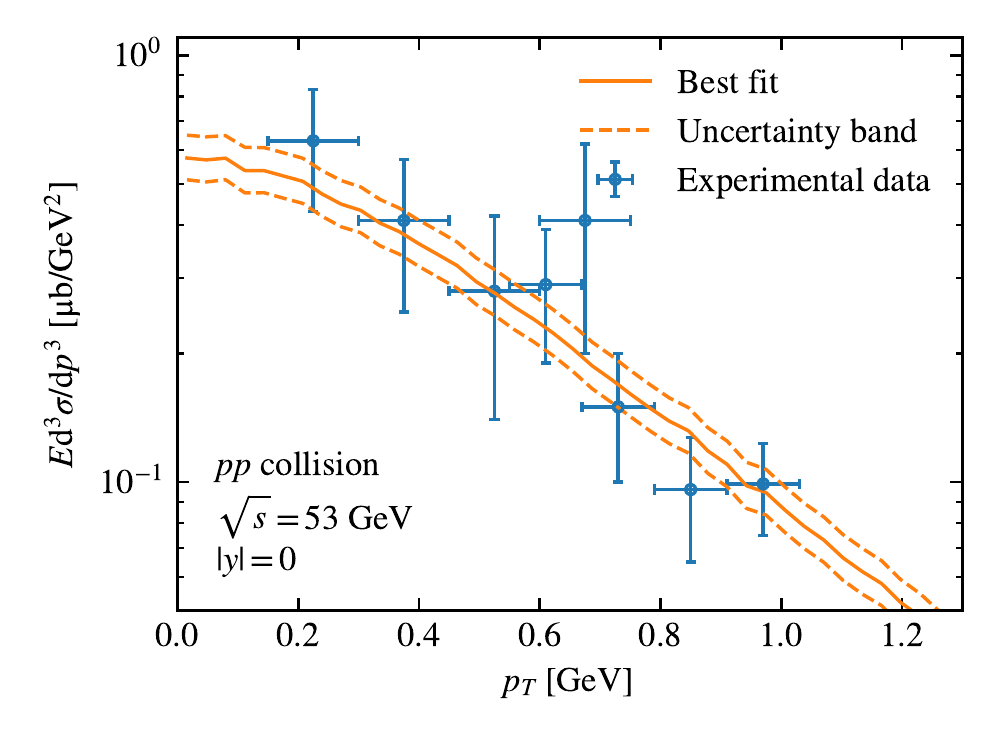}
   \includegraphics[width=0.49\columnwidth]{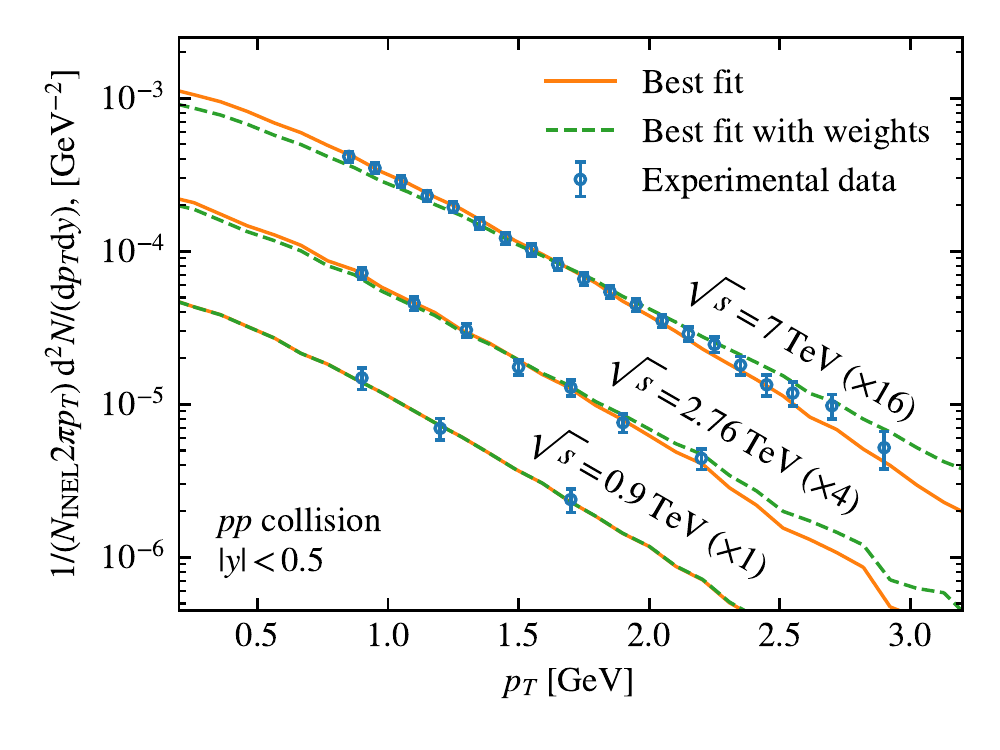}
    \vskip-0.3cm
    \caption{Best fit to antideuteron production data in $pp$ collisions, using
      QGSJET-II and the new coalescence model. Left: 
       results for  $\sqrt{s}=53$\,GeV, including an  uncertainty band,
       compared to CERN ISR data~\cite{Alper:1973my,Henning:1977mt}. 
       Right: calculations for LHC energies,  without and with the re-weighting
       to the $pp$ data on antiproton production, as discussed in the text,
       compared to ALICE data~\cite{collaboration_production_2018}.}
    \label{fig:pp}
\end{figure}

\begin{table}[htb]
    \centering
    \begin{tabular}{@{}lllllllll@{}}
    \toprule
                     & \multicolumn{2}{c}{LT $\zeta$} & \multicolumn{2}{c}{const. $\zeta$} &  \multicolumn{2}{c}{standard coal.}\\
        Experiment   & $\sigma$ [fm]   & $\chi^2/(N-1)$& $\sigma$ [fm]   & $\chi^2/(N-1)$ & $p_0$ [MeV]   & $\chi^2/(N-1)$\\ 
    \midrule
        p-p 7 TeV    & $1.44\pm0.01$  & 10/19 & $1.23\pm0.01$ & 86/19 & 134 & 93/19\\
        p-p 2.76 TeV & $1.29\pm0.03$ & 2.1/6 & $1.11\pm0.02$ & 9.9/6 & 146 & 12/6\\
        p-p 900 GeV  & $1.02\pm0.05$ & 0.30/2 & $0.90\pm0.04$ & 0.68/2 & 175 & 0.88/2\\
        p-p 53 GeV   & $0.50\pm0.03$ & 3.2/7 & $0.47\pm0.03$ & 2.9/7 & 280 & 2.5/7\\
        p-Be & $1.00\pm 0.03$  & 2.2/4 & $0.95\pm$0.03 & 2.4/4 &  126 & 3.0/4\\
        p-Al & $0.88\pm 0.04$  & 1.4/2 & $0.84\pm$0.04 & 1.5/2 &  126 & 1.3/2\\
    \bottomrule
    \end{tabular}
    \caption{Calibration results for antideuteron production, obtained using
      QGSJET-II: including the effect of the Lorentz transformation 
    on $\sigma_{\perp}$ [Eq.~(\ref{LT})], using constant $\sigma_{\perp}$, 
    and employing the standard coalescence model.}
    \label{tab:res_qgsjet}
\end{table}

\begin{table}[htb]
    \centering
    \begin{tabular}{@{}lllllllll@{}}
    \toprule
                     & \multicolumn{2}{c}{LT $\zeta$} & \multicolumn{2}{c}{const. $\zeta$} &  \multicolumn{2}{c}{standard coal.}\\
        Experiment   & $\sigma$ [fm]   & $\chi^2/(N-1)$& $\sigma$ [fm]   & $\chi^2/(N-1)$ & $p_0$ [MeV]   & $\chi^2/(N-1)$\\ 
    \midrule
        p-p 7 TeV    & $1.07\pm0.01$  & 29/19 & $0.92\pm0.02$ & 133/19 & 176 & 177/19\\
        p-p 2.76 TeV & $1.05\pm0.02$ & 8.7/6 & $0.93\pm0.04$ & 32/6 & 174 & 45.6/6\\
        p-p 900 GeV  & $0.97\pm0.05$ & 2.6/2 & $0.87\pm0.07$ & 6.1/2 & 181 & 7.3/2\\
        p-p 53 GeV  & $1.03\pm0.06$ & 3.3/7 & $0.96\pm 0.06$ & 2.7/7 & 171 & 2.1/7\\
        ALEPH        & $1.04^{+0.20}_{-0.12}$ & - & $0.99^{+0.18}_{-0.12}$ & - & $214^{+21}_{-26}$ & -\\
        ALEPH+OPAL   & $1.15^{+0.27}_{-0.22}$ & 3.2/1 & $1.09^{+0.26}_{-0.22}$ & 3.2/1 & 201 & 3.2/1\\
    \bottomrule
    \end{tabular}
        \caption{Calibration results for antideuteron production,
	 obtained using PYTHIA 8.230.}
    \label{tab:res_PYTHIA}
\end{table}

\begin{figure}[htbp]
    \centering
      \includegraphics[scale=\scale]{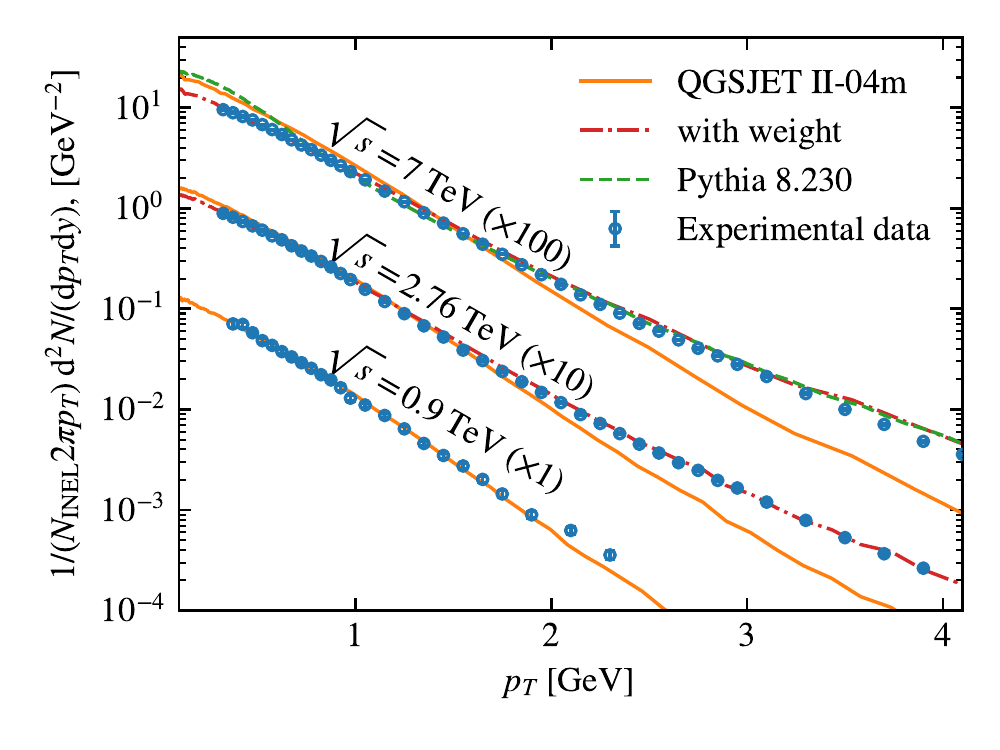}
    \vskip-0.45cm
    \caption{Combined invariant differential yield of protons and antiprotons
    in $pp$ collisions, obtained   using PYTHIA (green dashed line) and 
     QGSJET-II (solid orange lines), 
compared to   ALICE   data~\cite{Adam:2015qaa,Aamodt:2011zj,Abelev:2014laa}. 
    The red dashed-dotted lines are
    obtained by adding a multiplicative
    weight $w=ap_T^b + c$ to the 
    yield predicted by QGSJET-II.}
    \label{fig:pbarp_ALICe}
\end{figure}

Taking these results at face-value, one might interpret, e.g., the change
from $\sigma\simeq 0.5$\,fm at 53\,GeV to $\sigma\simeq 1.44$\,fm  at 7\,TeV,
using QGSJET-II, as an energy dependence of this parameter. However, such a
change may also be caused by a systematic bias either in the experimental data
and/or in the predictions of the used event generators. In order to clarify
the reason for this change, we compare in Fig.~\ref{fig:pbarp_ALICe} the
invariant differential yield of protons and antiprotons, measured by the ALICE
collaboration~\cite{Adam:2015qaa,Aamodt:2011zj,Abelev:2014laa}, to the values
obtained using QGSJET-II at $\sqrt{s}=900$, 2760 and 7000\,GeV, and using
PYTHIA at
$\sqrt{s}=7000$\,GeV. As is easily seen in the figure, QGSJET-II fits the
data at 900\,GeV well, but overestimates the bulk of the produced antiprotons
at 2760 and 7000\,GeV. Therefore, the coalescence parameter $\sigma$
must be artificially
increased at these energies to compensate the overproduction of
antinucleons. In the same manner, QGSJET-II underestimates the antiproton
flux measured at the CERN~ISR\footnote{A short
  discussion of this effect on the standard coalescence model and CERN~ISR data
  using PYTHIA and EPOS-LHC can be found in Ref.~\cite{ALICE-PUBLIC-2019-006}.}.
Thus, $\sigma$ has to be decreased for QGSJET-II to compensate this deviation.
In all the aforementioned cases, the deviations from the expected value
$\sigma\simeq 1$\,fm are caused by an imperfect description of antiproton
production by the Monte Carlo event generators.

In order to quantify this effect, we tweak the antiproton spectra by adding
a weight $w=ap_T^b + c$ and fix $a$, $b$ and $c$ by fits to the experimental
data. This implies that the weight $w_{\bar{d}}=[a(p_T/2)^b + c]^2$ has to be
included in the case of antideuteron production. We fit the weight $w$ to the
combined antiproton and proton data measured by
ALICE~\cite{Adam:2015qaa,Aamodt:2011zj,Abelev:2014laa}, with the same
experimental set-up as the antideuteron data, and to the antiproton data
measured
at the CERN~ISR\footnote{In the fit of the CERN ISR data, $b=1$ was fixed and
  only the data points in the interval $0.2\leq p_T\leq 1.3\unit{GeV}$ were
  used.}~\cite{Alper:1975jm}.
The resulting best-fit yields shown in
Fig.~\ref{fig:pbarp_ALICe} reproduce nicely the experimental data.
Then we repeat the analysis of the antideuteron data of ALICE and CERN ISR:
The values of $\sigma$ obtained using the re-weighted antinucleon spectra are
listed in Table~\ref{tab:res_qgsjet_update} and the fits to the 
antideuteron  data of ALICE are plotted in the right panel of
Fig.~\ref{fig:pp}.  In all the cases, the results are
significantly closer to the expected value $\sigma\simeq 1\unit{fm}$.
Note that the weights are specific for each experimental set-up
and kinematic  range: They were chosen to affect mainly the shape of the
antiproton spectra in the narrow kinematic range covered
by experimental data. In contrast, the total yields important for
astrophysical applications are less sensitive to the systematic
uncertainties of the event generator at large $p_T$.
For instance, the total antideuteron yield in $pp$ collisions
at $7\,$TeV, using $\sigma\simeq 1.1$\,fm,  would be
decreased by $\simeq 40\%$, relative to the case of using no weights.

Finally, let us compare our results to those of Ref.~\cite{Gomez-Coral:2018yuk}.
Imposing the coalescence condition in momentum space on an event-by-event
basis, the authors of that work used EPOS-LHC to reproduce experimental data
on the (anti)deuteron yield in $pp$ and $pA$ collisions. Based on these
comparisons, they suggested that $p_0$ is strongly energy dependent at low
energies\footnote{Note that their fit of $p_0$ as function of $p_{\rm lab}$
  combines data from $pp$ and $pA$  collisions, which correspond to
  different cm energies. Moreover, the differences in the size of the
  formation region
  of deuterons as function of $A$ are neglected in such an approach.}.
Moreover, they proposed that the energy dependence differs for deuteron and
antideuteron production:  While $p_0$ increases for deuterons, it decreases
for antideuterons as the kinetic energy of the projectile decreases. Such a
behaviour is difficult to understand,
if one accepts that the strong interaction does not distinguish between
matter and antimatter. In contrast, a possible contamination by deuterons
produced in the detector may easily explain the larger value of $p_0$ for
deuterons than for antideuterons.
From a theoretical point of view, we expect that the size of the formation
region---and thus $\sigma$---is only logarithmically dependent on the cm
energy. Furthermore, its size should be identical for deuteron and antideuteron
production. However, we have seen that a relatively small bias in the
production spectra of antinucleons or, alternatively, systematic errors in
the experimental results may delude an energy dependence of $\sigma$.
Correcting for such biases, we have verified that the present
experimental data
are consistent with the universal coalescence picture implemented in our
model. 
However, we note that the old data at $p_\mathrm{lab}=70\unit{GeV}$
from Serpukhov~\cite{Abramov:1986ti} are inconsistent with  this
picture. A confirmation of these data could falsify the assumption 
underlying our model.

Based on the best fit values after re-weighting, we fix for the following
analyses $\sigma=(1.0\pm0.1)$\,fm for both PYTHIA and QGSJET~II.
This value describes in our model via Eqs.~(\ref{w}--\ref{sigdef}) the
formation of light antinuclei for all interaction types and energies. For
comparisons, we set in the standard coalescence model $p_0=180$\,MeV for
proton-proton, proton-nucleus and nucleus-nucleus collisions, while we use
$p_0=210$\,MeV in DM annihilations.

\begin{table}[htb]
    \centering
    \begin{tabular}{@{}lllllllll@{}}
    \toprule
                     & \multicolumn{2}{c}{LT $\zeta$} & \multicolumn{2}{c}{const. $\zeta$}&  \multicolumn{2}{c}{standard coal.} \\
        Experiment   & $\sigma$ [fm]   & $\chi^2/(N-1)$& $\sigma$ [fm]   & $\chi^2/(N-1)$& $p_0$ [MeV]   & $\chi^2/(N-1)$\\ 
    \midrule
        p-p 7 TeV    & $1.17\pm0.01$  & 19/19 & $0.97\pm0.01$ & 16.2/19 & 165 & 23/19\\
        p-p 2.76 TeV & $1.16\pm0.02$ & 3.6/6 & $0.99\pm0.02$ & 4.1/6 & 161 & 5.7/6\\
        p-p 900 GeV  & $1.01\pm0.05$ & 0.30/2 & $0.89\pm0.04$ & 0.60/2 & 178 & 0.81/2\\
        p-p 53 GeV  & $0.94\pm0.06$ & 2.7/7 & $0.89\pm0.05$ & 3.2/7 & 170 & 4.3/7\\
    \bottomrule
    \end{tabular}
    \caption{Calibration results for antideuteron production in $pp$ collisions, obtained by using 
    QGSJET-II and applying an additional multiplicative weight  $a p_T^b + c$ to the predicted
    antinucleon yield, as discussed in the  text.}
    \label{tab:res_qgsjet_update}
\end{table}

\section{Antinucleus source spectra}
\label{sec:source_spectra}

\subsection{Secondary production}
\label{sec:secondary_production}

\begin{figure}[htbp]
    \centering
      \includegraphics[scale=\scale]{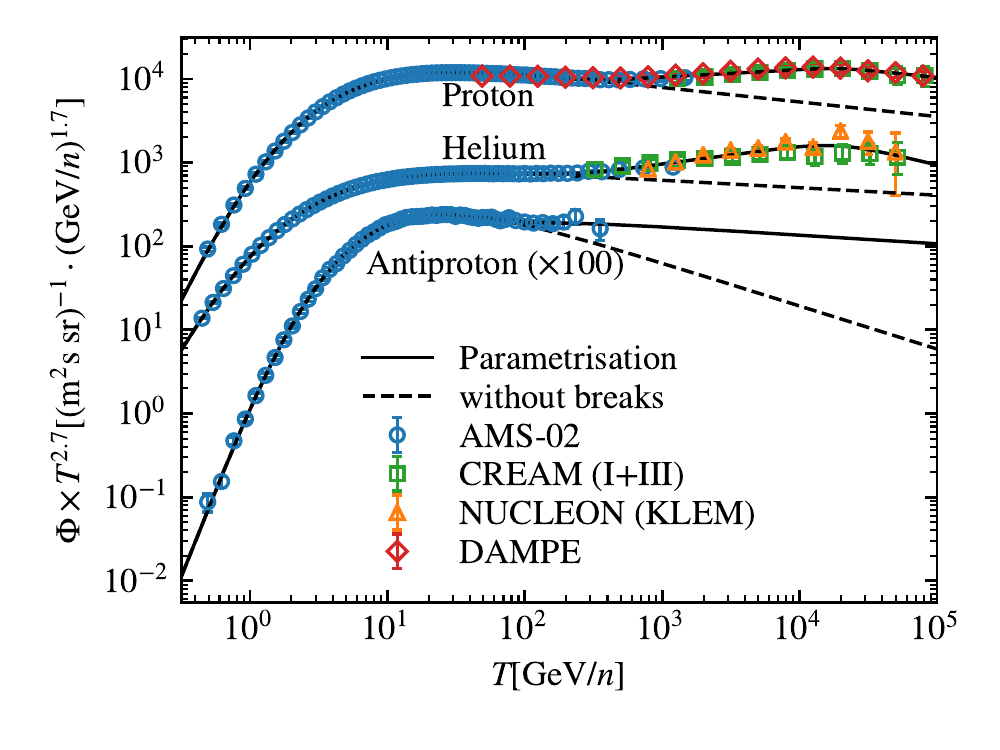}
    \vskip-0.4cm
    \caption{Parametrisations of the primary proton, helium and antiproton fluxes,
     compared to the data from 
     AMS-02~\cite{Aguilar:2015ooa,Aguilar:2015ctt,aguilar_antiproton_2016}, 
     DAMPE~\cite{An:2019wcw} and CREAM~\cite{Yoon:2017qjx}.}
    \label{fig:parametrisation}
\end{figure}

Light antinuclei are produced as secondaries in collisions of CRs with gas
in the Galactic disc. We neglect elements heavier than helium but take into
account the CR antiproton flux. The source term $Q^\mathrm{sec}$ of secondaries
can then be written as
\begin{equation}
    Q^\mathrm{sec}(T_{\bar{N}}, \vec r) = \sum_{i\in\{p, \mathrm{He}, \bar{p}\}}
    \sum_{j\in \{p, \mathrm{He}\}} 4\pi n_j(\vec r) \int_{T_{\bar{N}, \mathrm{min}}^{(i, j)}}^\infty\dd{T_i} \dv{\sigma_{i, j}(T_i, T_{\bar{N}})}{T_{\bar{N}}}\Phi_i(T_i, \vec r),
    \label{eq:sec_source}
\end{equation}
where $n_j(\vec r)$ is the density of particle $j$ in the Galactic disc,
$T=(E-m)/n$ is the kinetic energy per nucleon of the particle $i$ with mass
$m=nm_{\mathrm N}$ and flux $\Phi_i$, while $T_{\bar{N}, \mathrm{min}}^{(i, j)}$
is the threshold for creating an antinucleus $\bar{N}$.
We use as hydrogen density $n_\mathrm{H}=1\unit{cm^{-3}}$, while the helium
density is fixed to $n_\mathrm{He}=0.07n_\mathrm{H}$.
The differential cross section for $ij\to \bar{N}X$ is calculated as
\begin{equation}
    \dv{\sigma_{i, j}(T_i, T_{\bar{N}})}{T_{\bar{N}}}=\sigma_{ij,\mathrm{inel}}\dv{N_{\bar{N}}(T_i, T_{\bar{N}})}{T_{\bar{N}}},
\end{equation}
where $\sigma_{ij,\mathrm{inel}}$ is the total inelastic cross section, while
$\dv*{N_{\bar{N}}(T_i, T_{\bar{N}})}{T_{\bar{N}}}$ is computed using our
coalescence model. The parametrisations for the primary fluxes
$\Phi_i(T_i, \vec r)$ used in this work are compared
to experimental data in Fig.~\ref{fig:parametrisation}. We will employ two
parametrisations, one with and one without spectral breaks; their details are
discussed in Appendix~\ref{app:parametrisation}.

\begin{figure}[htbp]
    \centering
      \includegraphics[scale=\scale]{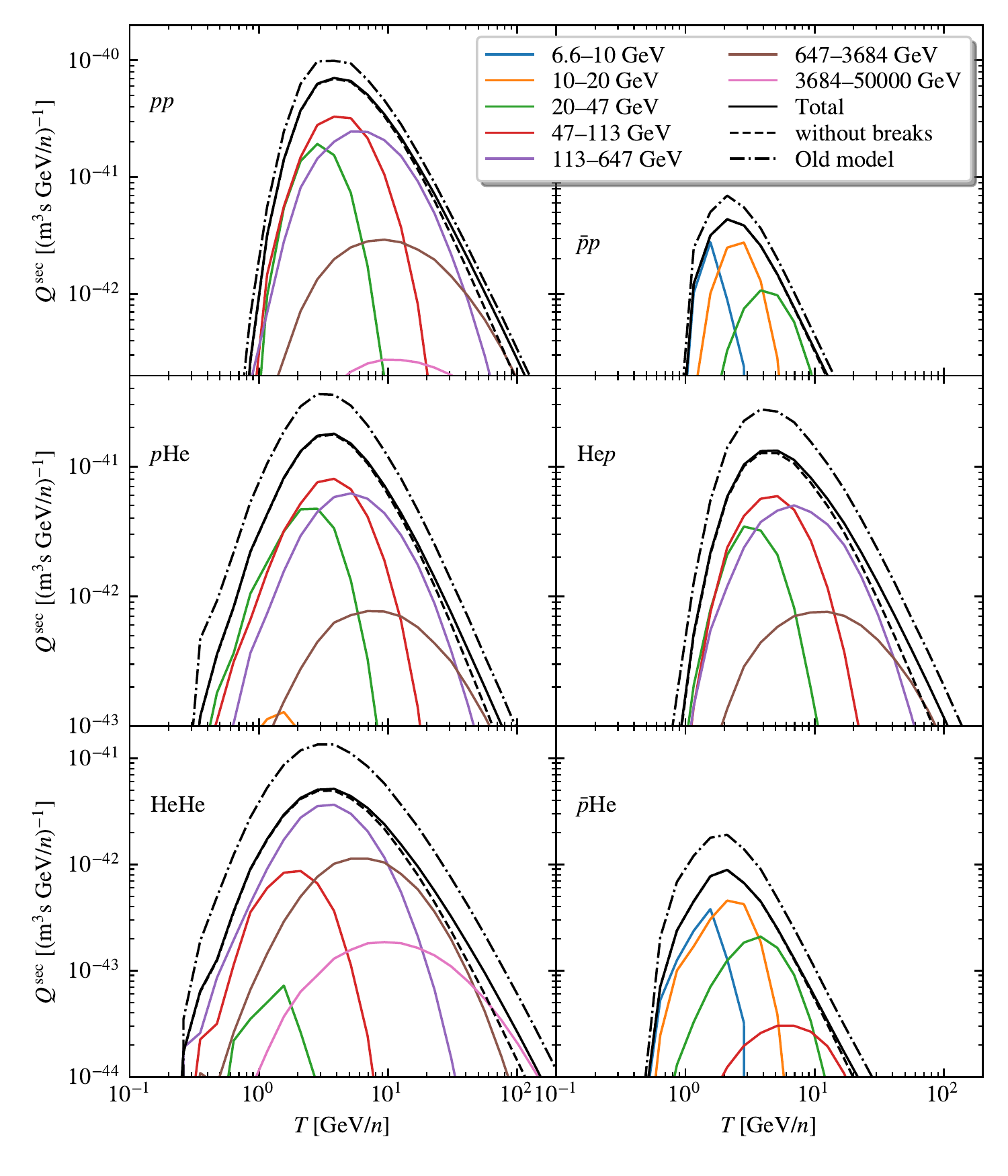}
    \caption{Partial contributions to the secondary source spectrum
      $Q^\mathrm{sec}$ of  antideuterons for the six different reactions
       and from various energy ranges, for the new model and primary
      spectra with breaks. Additionally, the resulting total contributions
      (black solid lines) are compared to the ones 
      obtained using the unbroken primary spectra (black dashed lines) 
      and the old coalescence model (dashed-dotted black lines). 
      The indicated energy ranges refer to the total energy 
      per nucleon for the He$p$ contribution and to the total energy
      of the primary particle in all the other cases.
    \label{fig:secondary_source_QGSJET_all}}
\end{figure}

We compute $\dv*{\sigma_{ij}}{T_{\bar{N}}}$ for 100~logarithmically spaced
energies $E_i$ of the projectile up to $5\times 10^4$\,GeV for
$i\in\{p, \mathrm{He},
\bar{p}\}$ and $j\in\{p, \mathrm{He}\}$. For each channel, we choose the
lower end of the energy range for $T_i$ such that all energies in which more
than $10^{-9}$ antideuterons per event are produced are included. The
contributions  of all these processes, for different incoming energy
ranges, are shown in Fig.~\ref{fig:secondary_source_QGSJET_all}. The
differences caused by the breaks in the primary spectra are negligible below
10 GeV/$n$ and small at higher energies; the largest difference appears for
the contribution from primary helium. Furthermore, the difference between
the new and standard
coalescence models is small in $pp$ and $\bar{p}p$ collisions, since
the parameter $p_0$ is adjusted to reproduce the correct yield of
antideuterons in $pp$ collisions. However, the  differences for the reactions
involving helium are larger, up to a factor $\sim 2$--3: While the new model
takes into account the increase in the size of the formation region of
antinucleons for helium, this effect is neglected in the case of the old
coalescence model. Therefore the old treatment tends to over-predict
the antideuteron yield in reactions involving helium.

\begin{figure}[htbp]
    \centering
      \includegraphics[scale=\scale]{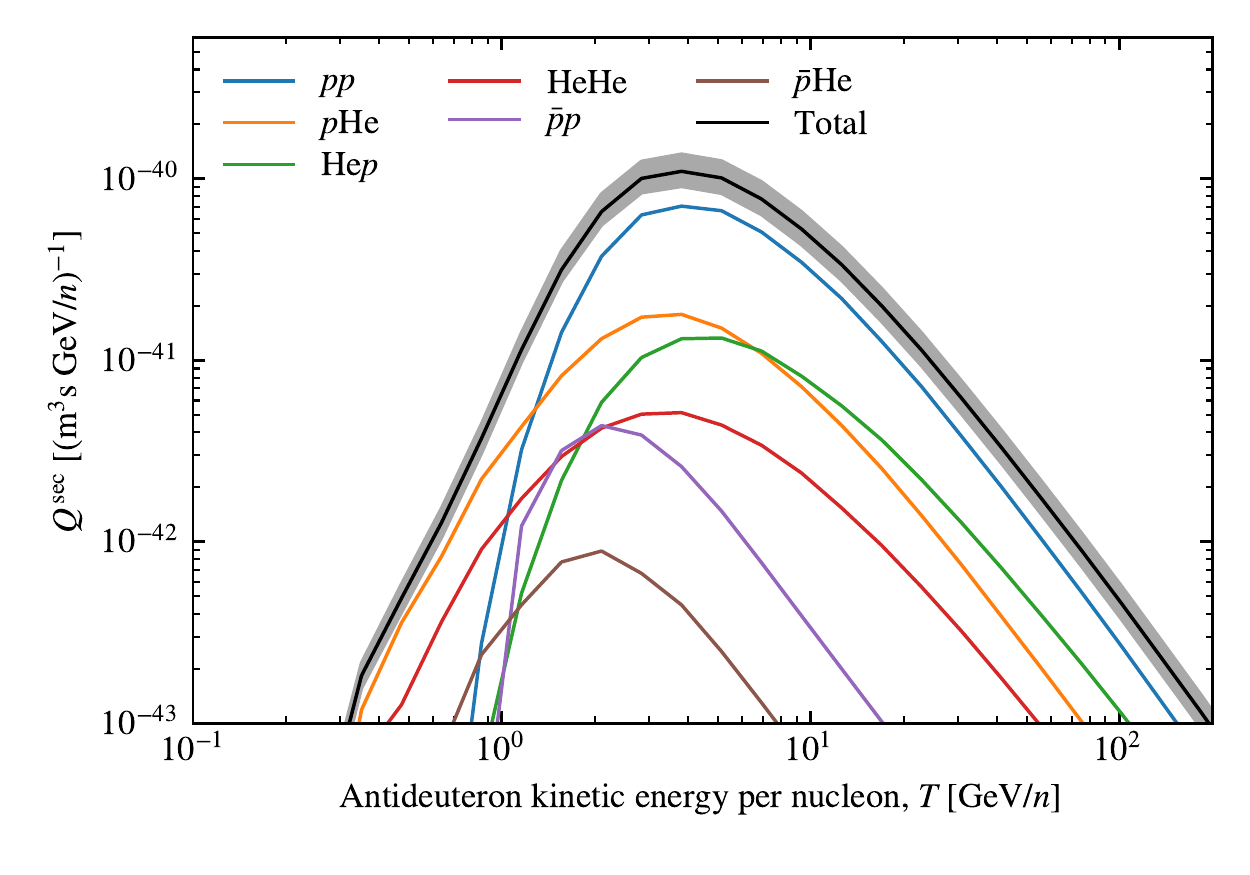}
    \vskip-0.45cm
    \caption{Contributions from different reaction types to the total secondary source term. The shaded area around the total corresponds to the estimated model uncertainty.}
    \label{fig:contributions_qgsjet}
\end{figure}

The contributions of the different reactions to the total secondary source
spectrum $Q^\mathrm{sec}$ of antideuterons are shown in
Fig.~\ref{fig:contributions_qgsjet}. Our results can
be compared to those of Lin {\it et al.}~\cite{Lin:2018avl} and Ibarra and
Wild~\cite{ibarra_prospects_2013}. Both groups used the standard coalescence
model in a Monte Carlo approach: Lin {\it et al.\/} employed the event
generators QGSJET-II-04m, EPOS-LHC and EPOS-1.99, while Ibarra and Wild used 
DPMJET-III and modified its results by adding a parametrised weight to the 
calculated antiproton spectra, in order to reproduce experimental data at low 
energies.
We find that the main contribution to the secondary source term comes from
$pp$ collisions, as expected. However, the low energy part is dominated 
by  $p$He and HeHe interactions, which is a consequence of the kinematics of
antideuteron production in these  reactions, in particular, of their lower
energy thresholds\footnote{In a proton-nucleus
collision, $\bar pp$ and $\bar nn$ pairs may be produced by partial inelastic
re-scatterings of the incident proton off two different  target nucleons. 
In a nucleus-nucleus scattering, on the other hand, such pairs may be produced 
by partial inelastic interactions between two different pairs of the projectile
and target nucleons.}: $T_{\bar{d}, \mathrm{min}}^{(p, {\rm He})}= 10m_N$ and 
$T_{\bar{d}, \mathrm{min}}^{({\rm He}, {\rm He})}= 6m_N$.
These findings are in contrast to the results of both Lin
{\it et al.\/} and Ibarra and Wild. Since both groups used
the so-called nuclear enhancement factor $\epsilon$, instead
of performing a proper calculation of the antideuteron production in $p$He,
He$p$ and $\bar{p}$He collisions, they could not observe this low-energy
behaviour.
The limitations of the concept of a nuclear enhancement factor $\epsilon$
were studied in some detail in Refs.~\cite{Kachelriess:2014mga,Kachelriess:2015wpa}.
In particular, the definition of $\epsilon $ assumes that the primary
CR fluxes are power laws without breaks. Moreover, the nuclear enhancement
factors for the production of massive particles are modified by threshold
effects and are thus strongly energy dependent in the energy range
relevant for astrophysical applications~\cite{Kachelriess:2015wpa}.

\begin{figure}[htbp]
    \centering
      \includegraphics[scale=\scale]{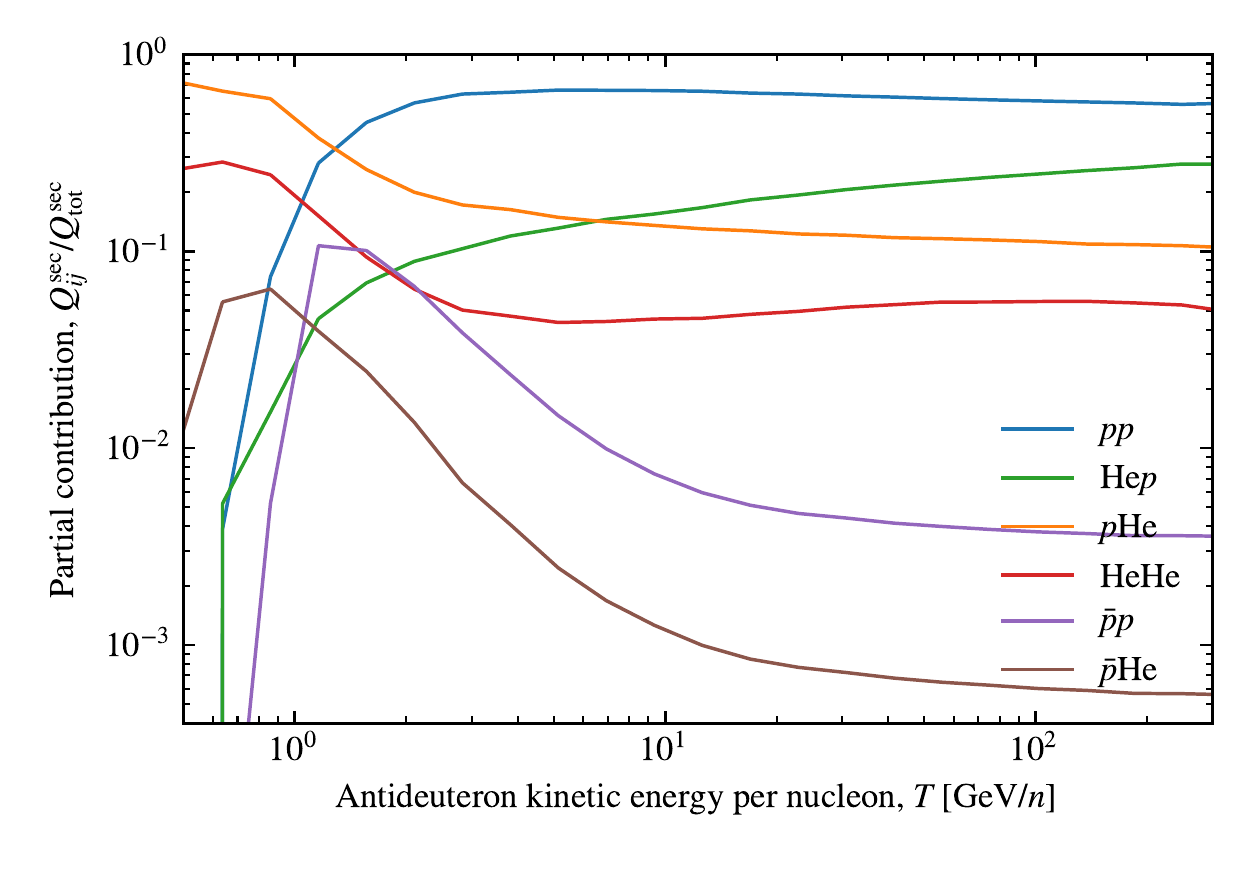}
    \vskip-0.45cm
    \caption{Relative contributions of the different reactions
      to the antideuteron source spectrum as function of kinetic energy
      per nucleon of the projectile.}
    \label{fig:eps}
\end{figure}

The contributions of different reaction types to the total secondary source
term are shown in Fig.~\ref{fig:contributions_qgsjet} for the new coalescence
model and the broken primary spectra. The shaded area around the total
contribution shown by the black solid line corresponds to the estimated model
uncertainty obtained by varying $\sigma$ in the range 0.9 to 1.1\,fm. As one
can see  in Fig.~\ref{fig:contributions_qgsjet}, the ``nuclear enhancement'',
i.e.\ the ratio of the values
corresponding to the black\footnote{Modulo the small contributions from 
$\bar{p}p$ and $\bar{p}$He collisions.}
 (total contribution) and blue ($pp$ contribution)
solid lines in the figure is indeed strongly energy-dependent.
This applies, in particular,  to the sub-GeV range where
the partial contributions to  the antideuteron source  term from  $p$He and
HeHe collisions exceed the one from proton-proton scattering by orders of
magnitude. The strong energy-dependence of the relative importance  of the
partial contributions to $Q^{\rm sec}$  at energies $T/n\lsim 10$\,GeV is
shown also clearly in Fig.~\ref{fig:eps}. We conclude from this figure that
the use of a nuclear enhancement factor at low energies should be avoided.

\begin{figure}[htbp]
    \centering
      \includegraphics[scale=\scale]{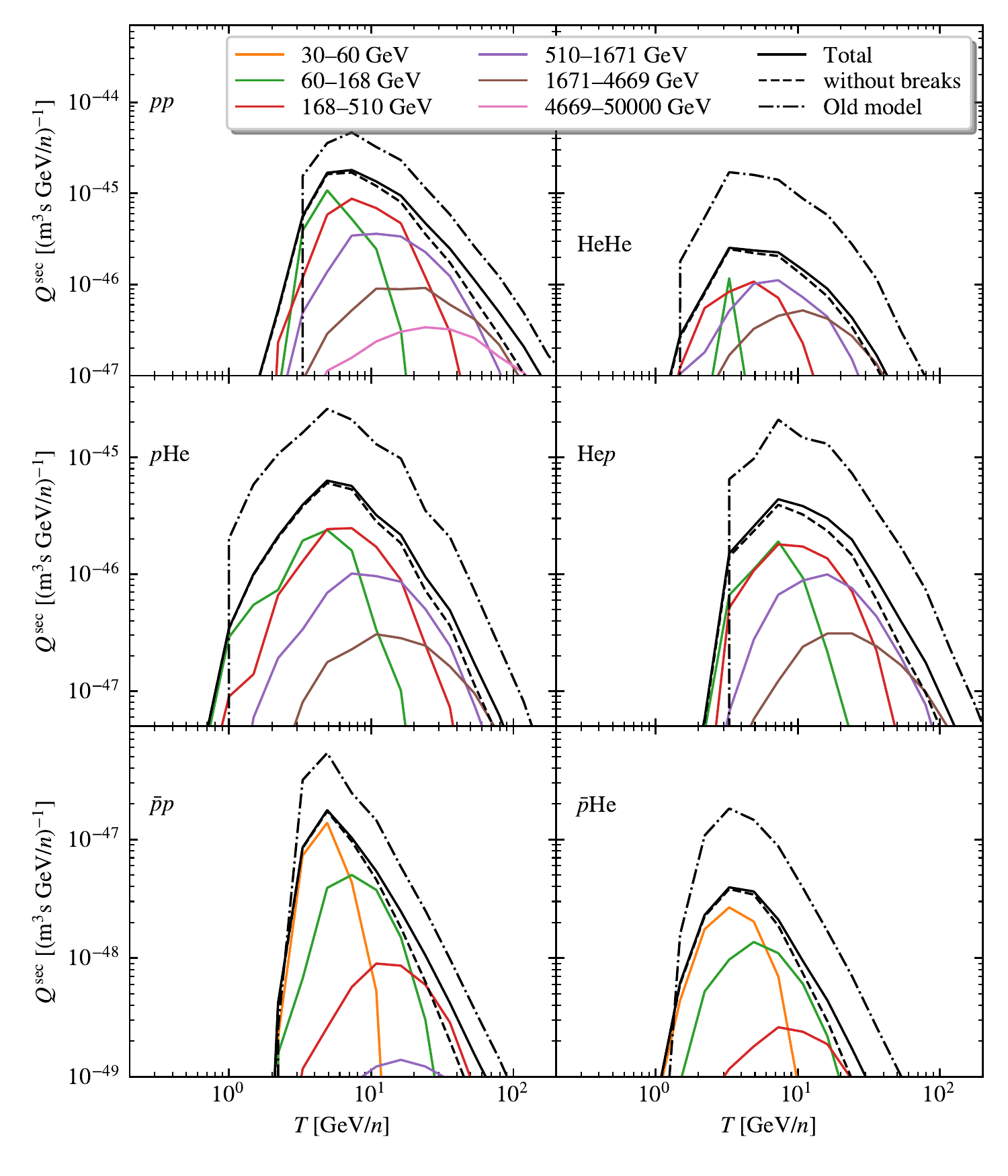}
    \caption{Partial contributions to the combined secondary source 
    spectrum $Q^\mathrm{sec}$ of antihelium and antitritium, similar to 
    Fig.~\ref{fig:contributions_qgsjet}.}
    \label{fig:helium_source_sec}
\end{figure}

\begin{figure}[htbp]
    \centering
      \includegraphics[scale=\scale]{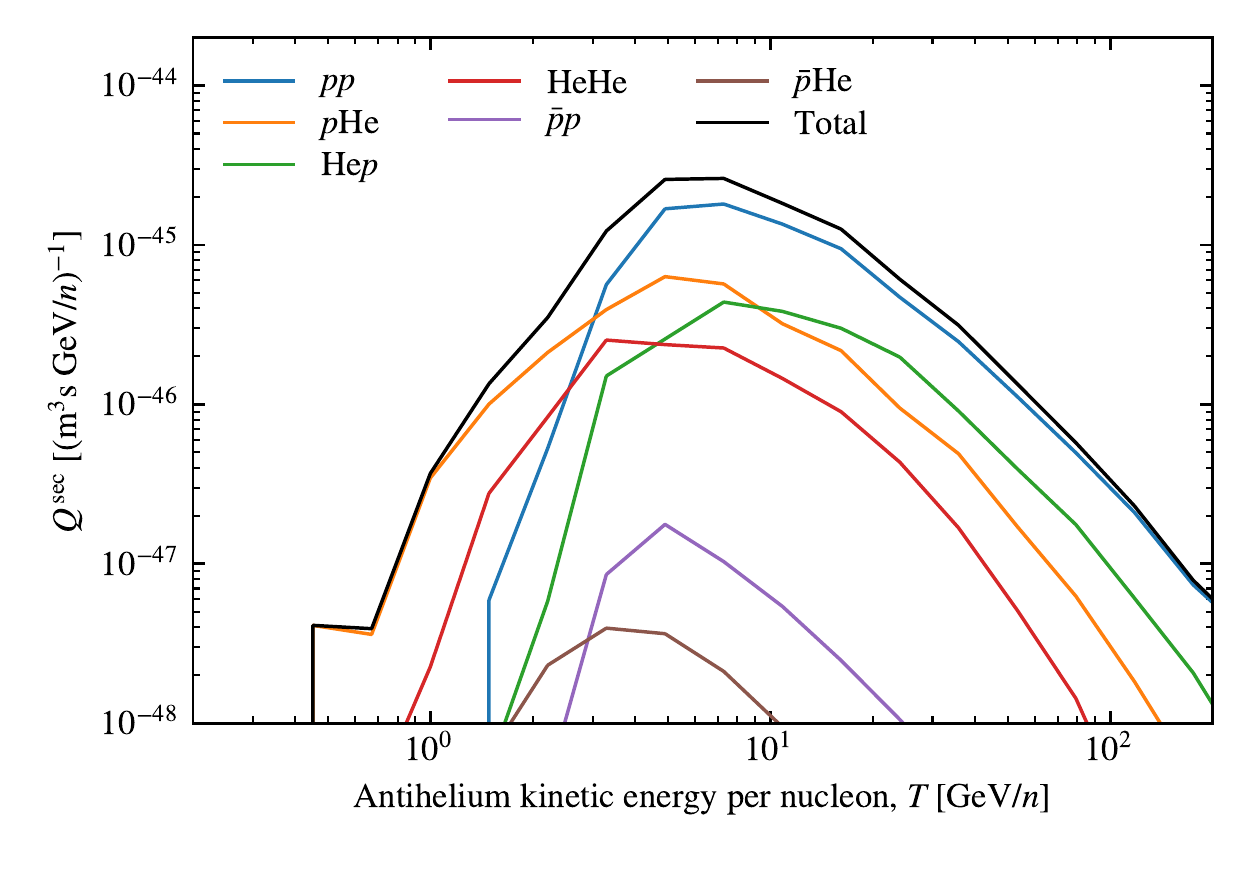}
    \vskip-0.45cm
    \caption{Contributions from different reaction types to the total secondary source term of antihelium.}
    \label{fig:contributions_qgsjet_helium}
\end{figure}

Following the same procedure, we have calculated antihelium production in $pp$,
$p$He, He$p$, HeHe, $\bar{p}p$ and $\bar{p}$He interactions, using
56~logarithmic bins from $E_p= 60$ to $5\times 10^3$\,GeV.
The resulting contributions to the source spectrum
are shown in Fig.~\ref{fig:helium_source_sec} for different energy ranges
of the primary particles. The relative contributions from the various
interactions are compared in Fig.~\ref{fig:contributions_qgsjet_helium}.
Since tritium decays fast compared to the propagation
time scale, the plotted antihelium source spectrum includes also the
antitritium contribution.

Comparing the source term of antihelium
to the one of antideuteron, we see that its maximum is reduced by a factor
$\mathrm{few}\times 10^4$ and shifted somewhat to larger $T/n$. 
Otherwise, the same qualitative features discussed above for the
antideuteron source term are still present. 
In particular, because of the possibility to interact with multiple
target nucleons simultaneously, the contribution to the source term 
from $p$He and HeHe collisions dominates the one from $pp$ interactions up to  
few GeV/nucleon.
There is also a notable difference concerning the relative contributions
of $\bar{p}p$ and  $\bar{p}$He collisions:
As expected from threshold effects,%
\footnote{
  Near threshold, an antideuteron is produced in the
  interactions $pp\to\bar{p}\bar{n}pppn$ and $\bar{p}p\to\bar{p}\bar{n}pn$,
  while an antihelium is produced in $pp\to\bar{p}\bar{p}\bar{n}ppppn$ and
  $\bar{p}p\to\bar{p}\bar{p}\bar{n}ppn$. Thus, there is a larger relative 
  change in going from antideuteron to antihelium in $\bar{p}p$ interactions
  than in $pp$.
}
those are up to a
factor $\sim10$ smaller, compared to the antideuteron case,
 at $T/n\lesssim 50\unit{GeV}$.
In any case, the contribution from $\bar{p}p$ interactions on the final
antinuclei spectra is neglible.

\subsection{Dark matter annihilations}

\begin{figure}[htbp]
    \centering
      \includegraphics[scale=\scale]{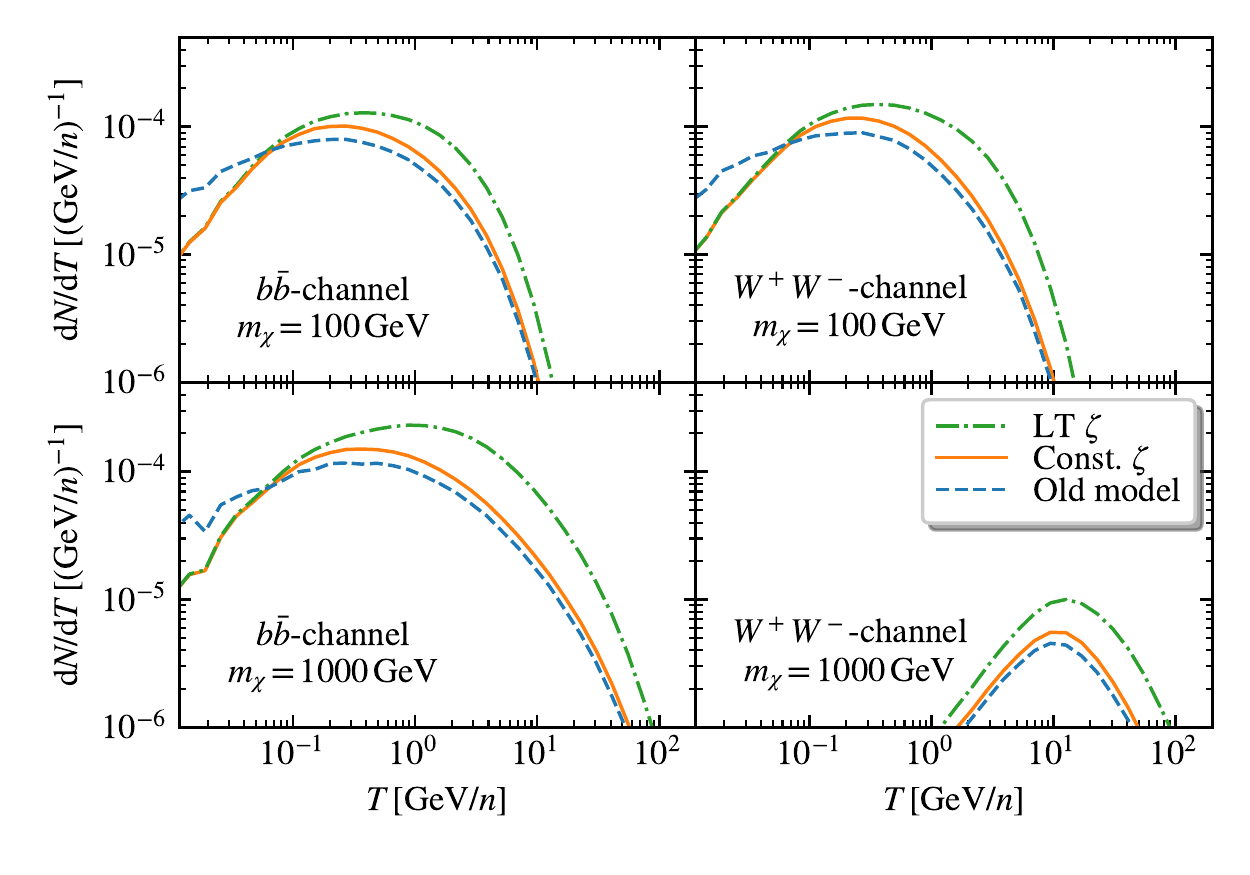}
    \caption{Antideuteron injection spectra from DM annihilations into
      $b\bar b$ (left) and $W^+W^-$ (right), for $m_\chi=100$\,GeV (top) and
      $m_\chi=1000$\,GeV (bottom).}
    \label{fig:DM_injection}
\end{figure}

\begin{figure}[htbp]
    \centering
       \includegraphics[scale=\scale]{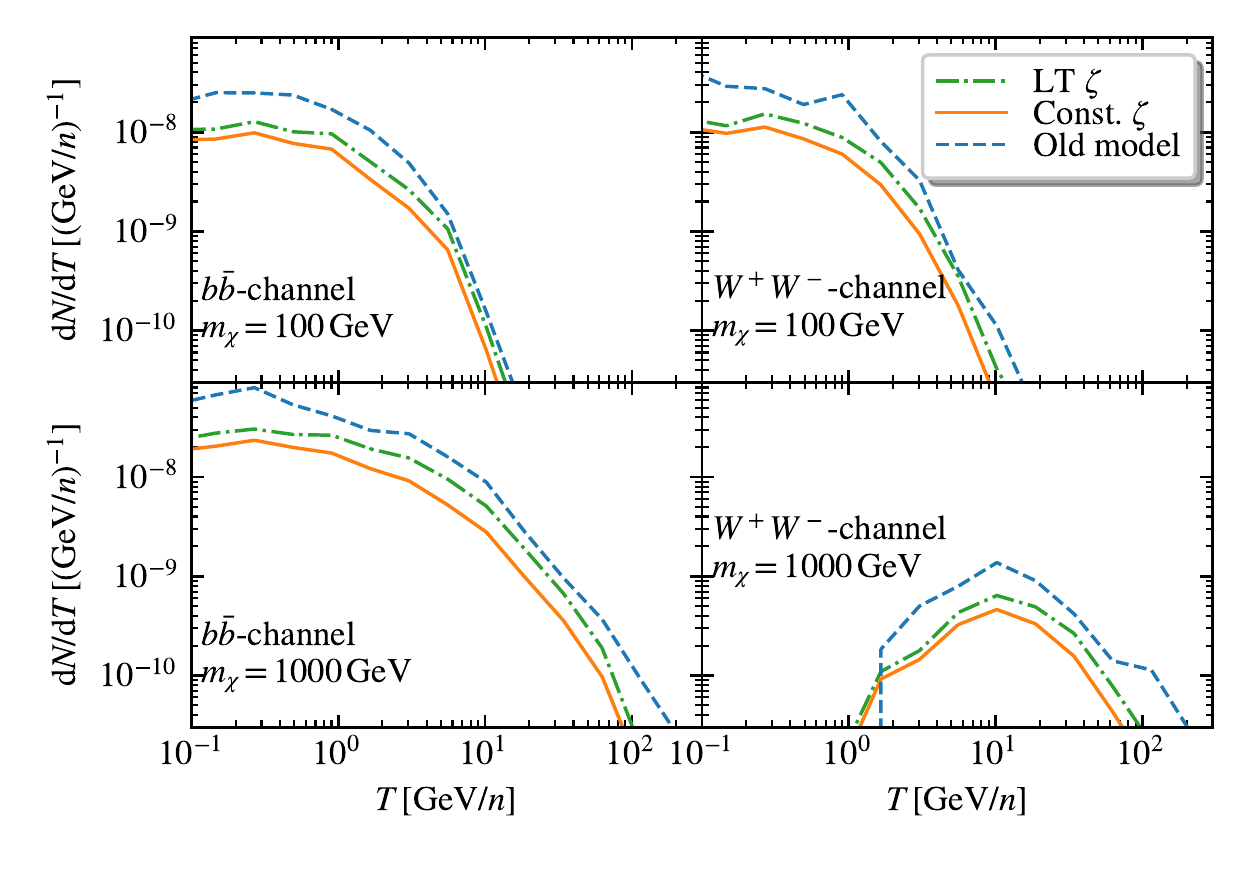}
    \caption{Antihelium (plus antitritium) injection spectra from
      DM annihilations into $b\bar b$ (left) and $W^+W^-$ (right), for
      $m_\chi=100$\,GeV (top) and $m_\chi=1000$\,GeV (bottom).}
    \label{fig:DM_injection_antihelium}
\end{figure}

In addition to the secondary production, we consider antinuclei originating
from DM annihilations. We consider as DM particles Majorana fermions which
annihilate purely into $\bar{b}b$ or $W^+W^-$ pairs. These annihilations will
be modelled in PYTHIA by generating a generic collision of a non-radiating
$e^+e^-$ pair with $\sqrt{s}=2m_\chi$.  The injection spectra
$\dv*{N}{T}$ are shown in Fig.~\ref{fig:DM_injection} for antideuterons
and in Fig.~\ref{fig:DM_injection_antihelium} for antihelium and antitritium.
In both cases, we consider 100 and 1000\,GeV  as DM mass.
Note that in the antideuteron injection spectra the differences between the
standard and our new coalescence model can reach a factor of few, while they
are much smaller in the spectra of antihelium-3. The reason for this mismatch
are the wave-functions of the two nuclei: Since the one of the antideuteron 
is stronger peaked at $r=0$ than the one of antihelium, large values of 
$q$ in Eq.~(\ref{w}) are less suppressed for antideuterons, cf.\ with Fig.~4 of
Ref.~\cite{Kachelriess:2019taq}. As a result, the differences in the shape of
the antideuteron energy spectrum are more pronounced compared to the case 
of antihelium.

The DM source spectrum can be written as~\cite{Jungman:1995df}
\begin{equation}
    Q(\vec r, T) = \frac{1}{2}\frac{\rho^2(\vec r)}{m_\chi^2}\left<\sigma_\mathrm{ann} v\right>\dv{N_{\bar{N}}^i}{T_{\bar{N}}},
\end{equation}
where $\rho(\vec r)$ is the DM mass density, $m_\chi$ its mass,
$\left<\sigma_\mathrm{ann} v\right>$ its thermally averaged annihilation
cross section and $\dv*{N_{\bar{N}}}{T_{\bar{N}}}$ is the differential
number density of the antinuclei $\bar{N}$. An upper bound on
$\left<\sigma_\mathrm{ann} v\right>$ will be determined in
Sec.~\ref{ssec:upper_bound}, requiring that the antiproton flux measured by
AMS-02
is not exceeded. The effect of different DM density profiles $\rho(\vec r)$
is small compared to the propagation uncertainty and has already been
extensively discussed, see e.g.~Refs.\cite{ibarra_prospects_2013,Fornengo:2013osa}. For simplicity, we will therefore only use an Einasto profile
with $\alpha=0.17$, $r_s=28.4$\,kpc and
$\rho_s=0.033 \unit{GeV/cm^3}$~\cite{cirelli_pppc_2011}.

\section{Antinuclei fluxes}
\label{sec:prop_fluxes}

\subsection{Propagation model}

Charged particles diffuse in the turbulent Galactic magnetic field. We employ
the two-zone diffusion
model~\cite{1969ocr..book.....G,1980Ap&SS..68..295G} to describe the
propagation of antinuclei through the Milky Way, which provides a simplistic
but rather successful description of a variety of CR data.
In this scheme, the Galaxy with radius $R=20$\,kpc is
modelled as a cylinder containing a large diffusive CR halo of half-height $L$ 
and a thin disk of half-height $h\ll L$. The latter comprises the CR
sources and the interstellar medium which serves as target for secondary
production. In this model, the diffusion equation for the differential number
density
$n_{\bar{N}}$ of antinuclei can be written in cylindrical coordinates
$\vec r=(r, z)$, where $z$ is the height above the Galactic plane, as
\begin{equation}
\begin{aligned}
    -K \nabla^2 n_{\bar{N}} + \mathrm{sign}(z) V_{\rm c}\partial_zn_{\bar{N}}
    + 2h\delta(z)\partial_E\left(b_\mathrm{loss}n_{\bar{N}} - D_{EE}\partial_E n_{\bar{N}}\right)\\
    = Q^\mathrm{prim} + 2h\delta(z)\left[Q^\mathrm{sec}+Q^\mathrm{ter} -
    \Gamma_\mathrm{ann}n_{\bar{N}}\right],
    \label{eq:two_zone_diffusion_equation}
\end{aligned}
\end{equation}
where we have taken the limit $h=100\,\mathrm{pc}\ll L$. 
We parametrise the rigidity-dependent diffusion coefficient as a simple
power law,
\begin{equation}
    K(\mathcal{R}) = \beta K_0 (\mathcal{R}/{\rm GV})^\delta,
\end{equation}
with $\mathcal{R}=E/(Ze)$, and  $K_0$ and $\delta$ are free parameters. In
turn, diffusion in momentum space, which is included as second-order
re-acceleration in Eq.~(\ref{eq:two_zone_diffusion_equation}), is related to
$K(\mathcal{R})$ by
\begin{equation}
  D_{EE}(\mathcal{R})= 
\frac{4}{3\delta(4-\delta^2)(4-\delta)}V_{\rm A}^2\frac{v_{\bar{N}}^2p_{\bar{N}}^2}{K(\mathcal{R})}\,,
\end{equation}
where $V_{\rm A}$ is the Alfvén velocity, $v_{\bar{N}}$ is the velocity of the
antinuclei and $p_{\bar{N}}$ their momentum. Moreover,
$V_{\rm c}$ denotes the convection velocity which is assumed to be constant and
directed away from the Galactic disc, while $\Gamma_{\mathrm{ann}}^{\bar{N}} $ is
the annihilation rate of the antinuclei. The factor $b_\mathrm{loss}$ accounts
for Coulomb, ionization and adiabatic energy losses.
The primary proton, helium and antiproton
fluxes are assumed to be the same in entire Galactic disc. The flux of
antinuclei is related to the number density by
$\Phi(E, \vec r)=v n_{\bar{N}}(E, \vec r)/(4\pi)$.

The interaction rate of an antinucleus $\bar{N}$ will be approximated by
$\Gamma_i^{\bar{N}p} = (n_\mathrm{H} + 4^{2/3}n_\mathrm{He}) v_{\bar{N}} \sigma_{\bar{N}p}^i$, where the factor $4^{2/3}$ accounts approximately for the cross section
difference between helium and hydrogen, and $\sigma_{\bar{N}p}^\mathrm{ann}$ is
the  $\bar{N}p$ annihilation cross section. For antiprotons and antideuterons,
we find the cross sections using the procedure discussed in
Ref.~\cite{delahaye_antideuterons_2015}, while for antihelium-3 we follow
Ref.~\cite{Carlson:2014ssa}.

The tertiary term can be written as
\begin{equation}
\begin{aligned}
    Q^\mathrm{ter}(T_{\bar{N}}, \vec r) &= 4\pi (n_\mathrm{H} + 4^{2/3}n_\mathrm{He}) \left[\int_{T_{\bar{N}}}^\infty \dd{T_{\bar{N}}'} \dv{\sigma^\text{non-ann}(\bar N(T_{\bar{N}}') + p\to \bar{N}(T_{\bar{N}})+ X)}{T_{\bar{N}}} \Phi_{\bar{N}}(T'_{\bar{N}}, \vec r)\right.\\
    &\left. -\sigma^\text{non-ann}(\bar N(T_{\bar{N}}) + p\to \bar{N}(T_{\bar{N}}^{\prime\prime})+ X)\vphantom{\int_{T_{\bar{N}}}^\infty}\Phi_{\bar{N}}(T_{\bar{N}}, \vec r)\right],
\end{aligned}
\end{equation}
where $\Phi_{\bar{N}}(T_{\bar{N}}, \vec r)$ is the antinucleus flux at energy
$T_{\bar{N}}$. Thus, the tertiary terms are themselves dependent on the antinucleus
flux and Eq.~\eqref{eq:two_zone_diffusion_equation} becomes an
integro-differential equation that we solve using the method presented
in Ref.~\cite{donato_antiprotons_2001}.

\begin{table}
    \caption{Parameters used for the two-zone propagation model.}
    \label{tab:two-zone_params}
    \centering
    \begin{tabular}{@{}lccccc@{}}
    \toprule
    Model & $L$ [kpc] & $\delta$ & $K_0$ [$\mathrm{kpc^2/Myr}$] & $V_{\rm c}$
    [$\mathrm{km/s}$] & $V_{\rm A}$ [km/s] \\ \midrule
    max & 15 & 0.46 & 0.0765 & 5 & 117.6 \\
    med & 4 & 0.7 & 0.0112 & 12 & 52.9 \\
    KRW & 13.7 & 0.408 & 0.0967 & 0.2 & 31.9 \\
    Kolmogorov & 5 & 1/3 & 0.018 & 0 & 0 \\
    \bottomrule
    \end{tabular}
\end{table}

For  antinuclei from WIMP annihilations, we neglect both re-acceleration and
energy losses in Eq.~\eqref{eq:two_zone_diffusion_equation}, 
as they have been shown to have little impact on the final primary spectrum
for $T\gtrsim 1\unit{GeV}$~\cite{maurin_transport_2006}.
We also neglect the tertiary contribution to the primary flux, since 
it is small for antiprotons and antideuterons because of their small
non-annihilating inelastic cross section. For comparison, neglecting the
tertiary contribution in the case of helium-3 leads to a flux that is
roughly 40\% lower compared to the opposite limit of neglecting the
non-annihilating interactions~\cite{Carlson:2014ssa}. In this case, one can
use the common semi-analytical solution detailed, e.g., in Refs.~\cite{maurin_transport_2006,cirelli_pppc_2011,Fornengo:2013osa}. Note, however, that parts
of the estimated sensitivities of the upcoming GAPS and AMS-02 experiments
fall within the region $T\lsim 1\unit{GeV}$~\cite{aramaki_antideuteron_2016},
which means that one should include the losses in a complete analysis of
the low-energy range.

The final propagation model depends on five parameters:
$L, K_0, \delta, V_{\rm c}$ and $V_{\rm A}$. In order to ease the comparison to
earlier works, we employ the two  parameter sets dubbed `med' and `max'
in Ref.~\cite{Donato:2008yx}.  In addition, we use one parameter set inspired
by a plain Kolmogorov diffusion model and the best-fit parameters from a
recent B/C analysis~\cite{Kappl:2015bqa} performed by
Kappl\footnote{Note that Kappl~{\it et al.} used
  $n_\mathrm{H}=0.9\unit{cm^{-3}}$  and $N_\mathrm{He}=0.1\unit{cm^{-3}}$,
  meaning that our results are 9\% higher as if we would use their parameter
  set.}~{\it et al.} For the former, we fix $K_0$
by requiring that the grammage $X=c\rho hL/(2K)$ crossed by CR protons with
energy 10\,GeV equals 10\,g/cm$^2$.
The numerical values of the five parameters determining the propagation
model are summarised in Table~\ref{tab:two-zone_params}. Finally, we account
for Solar modulations using the force-field
approximation~\cite{gleeson_solar_1968,moraal_cosmic-ray_2013} with a Fisk
potential $\phi=0.6$\,GV, as described in Appendix~\ref{app:parametrisation}.

\subsection{Upper bound on the annihilation cross section from AMS-02 antiproton data}
\label{ssec:upper_bound}

\begin{figure}[htbp]
    \centering
      \includegraphics[scale=\scale]{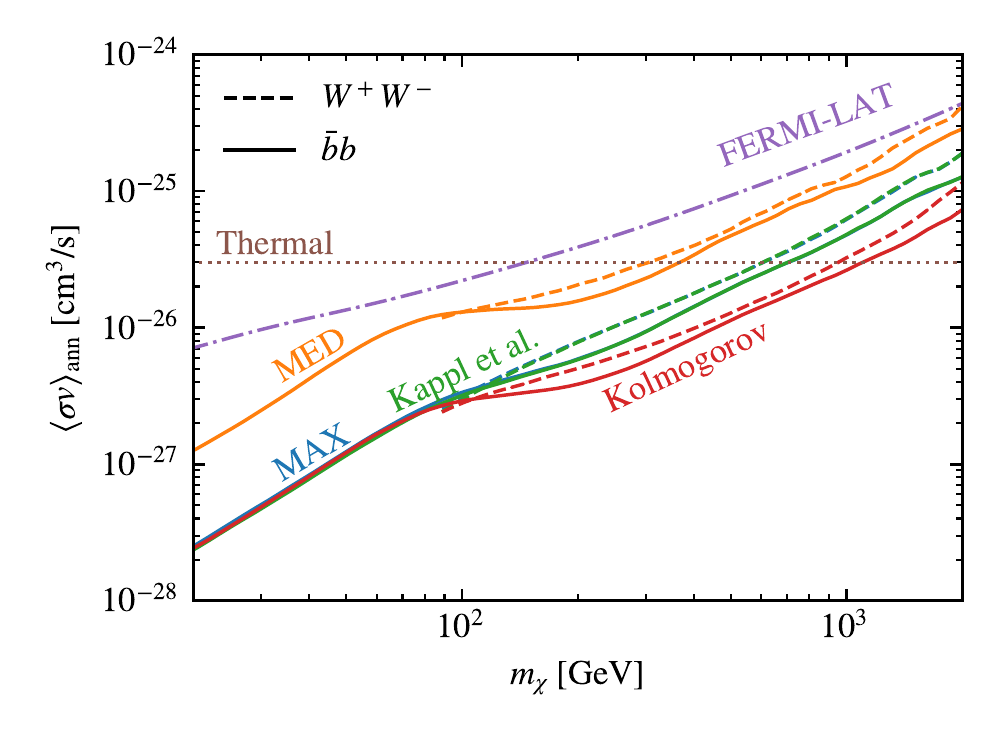}
    \caption{Upper limit compatible with the AMS-02 antiproton data
     for different propagation parameters. 
     The upper limit from Fermi-LAT~\cite{Ackermann:2015zua} (for $b\bar b$) 
     and the value $\left<\sigma_\mathrm{ann} v \right>$ favoured by cosmology
     are shown for comparison. }
    \label{fig:upper_limit}
\end{figure}

The generic DM model used in this work has, apart from the branching ratios,
two parameters: the DM mass $m_\chi$ and the thermally averaged annihilation
cross section
$\left<\sigma_\mathrm{ann} v \right>$. We will here investigate the maximal
flux of antinuclei consistent with the AMS-02 antiproton
data~\cite{Fornengo:2013osa}. There is
currently no clear evidence for an exotic primary component in the antiproton
spectrum  and we use this absence to set an upper bound on the annihilation
cross section for various DM masses. More precisely, we derive upper bounds
on $\left<\sigma_\mathrm{ann} v \right>$ by choosing as null-hypothesis the
fit to the antiproton flux shown in Fig.~\ref{fig:parametrisation}. We then
in turn vary the annihilation cross section until the $\chi^2$ value differs
by 3.84 from the null-hypothesis, corresponding to an 95\% CL upper
limit~\cite{patrignani_review_2016}.  A stringent upper bound compatible with
the antiproton flux is obtained when the same parameters are used in the
antideuteron and antihelium cases. The results are shown in
Fig.~\ref{fig:upper_limit} for the considered parameter sets and the
annihilation channels $W^+W^-$ and $\bar{b}b$. The canonical value for a thermal
relic, $\left<\sigma_\mathrm{ann} v \right>=3\times 10^{-26}\unit{cm^3/s}$, and
the upper bound obtained by the Fermi-LAT collaboration using
dwarf galaxies~\cite{Ackermann:2015zua} are plotted for comparison too.
Note that the antiproton limits are more stringent  than the recent
Fermi-LAT bound. However, we stress that these limits only hold for
the specific propagation model and astrophysical parameters used.

\subsection{Detection prospects}

\begin{figure}[htb]
    \centering
     \includegraphics[scale=\scale]{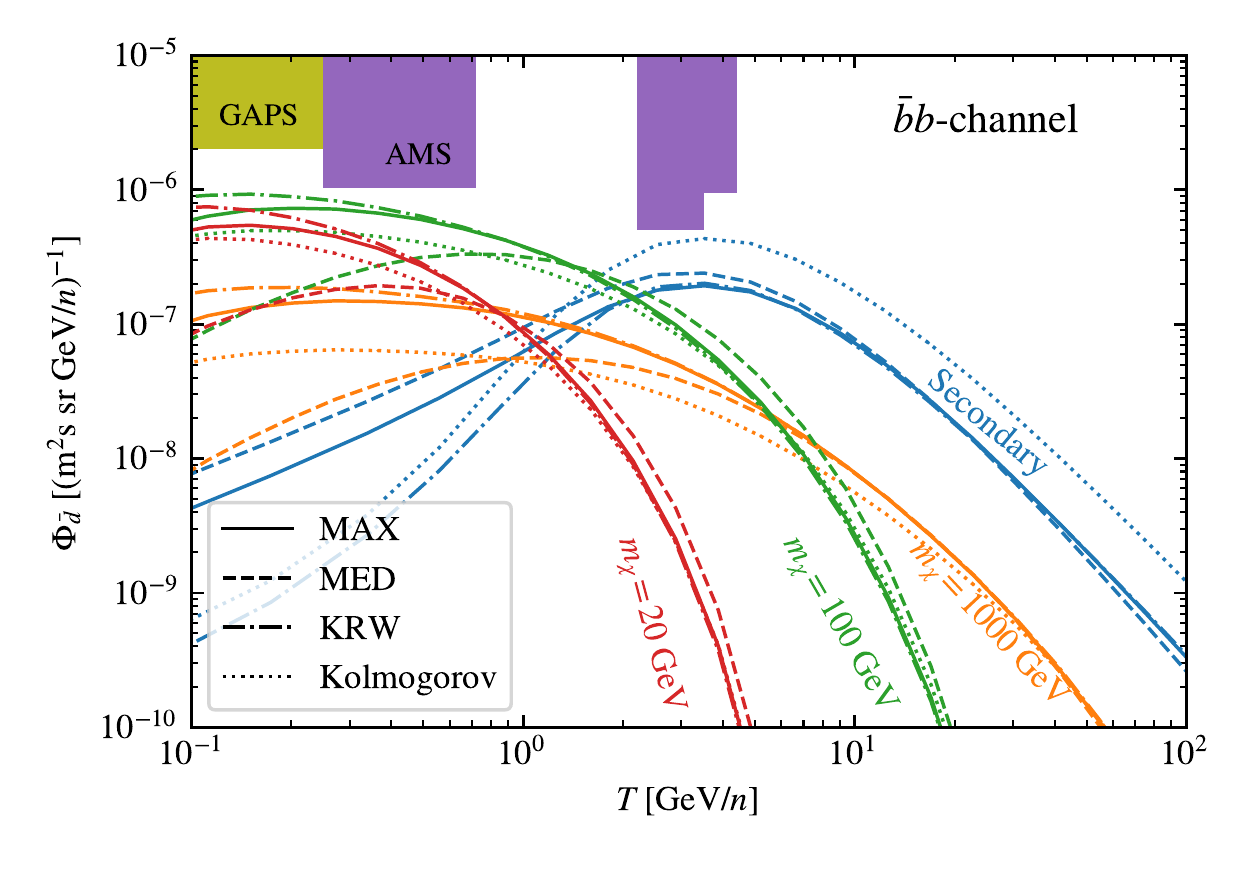}
    \caption{Estimated antideuteron flux on Earth from DM annihilations into $\bar bb$ pairs
    and from secondary production for the considered benchmark cases. 
    The shaded areas on the top are the estimated AMS-02 and GAPS sensitivities.}
    \label{fig:deuteron_flux_bb}
\end{figure}

\begin{figure}[htb]
    \centering
      \includegraphics[scale=\scale]{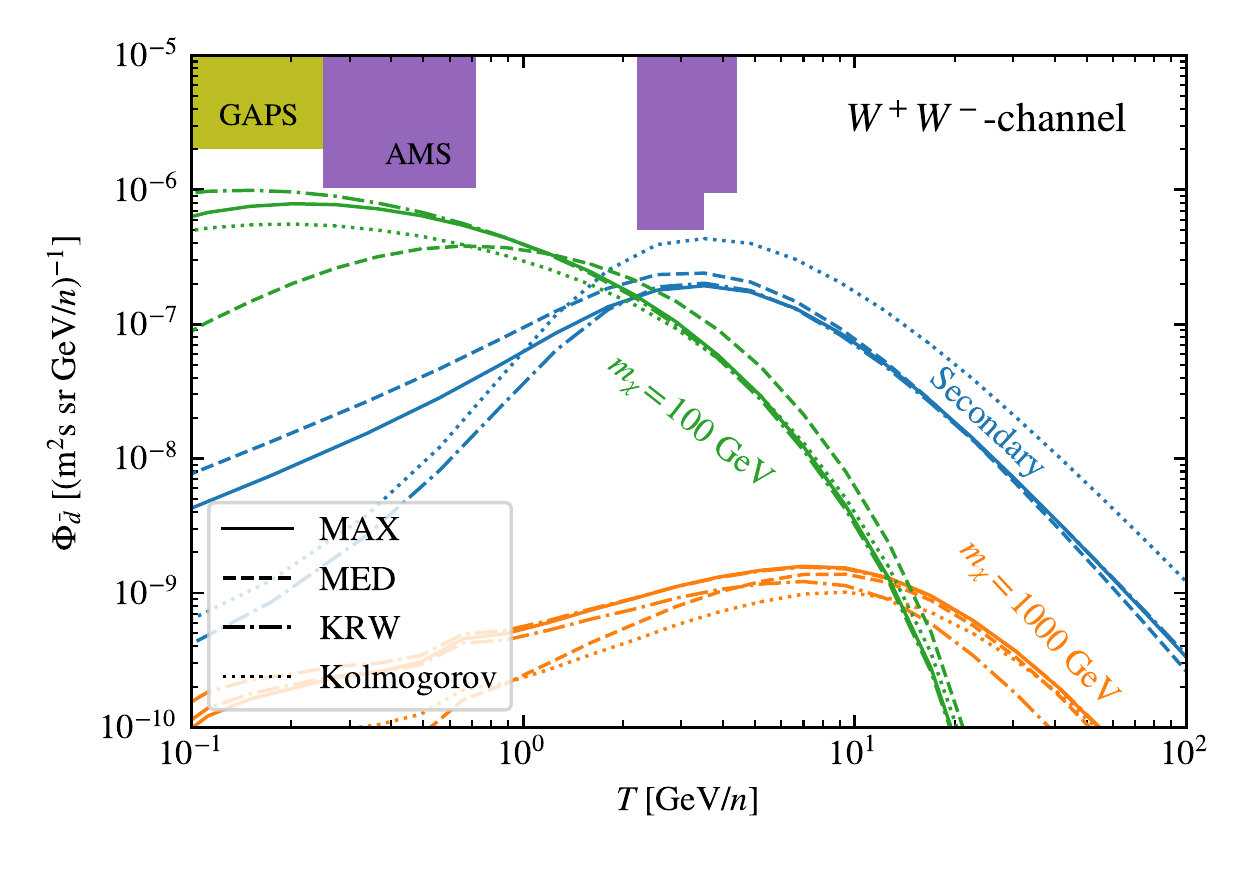}
    \caption{Estimated antideuteron flux on Earth from DM annihilations into $W^+W^-$ pairs
    and from secondary production for the considered benchmark cases. 
    The shaded areas on the top are the estimated AMS-02 and GAPS sensitivities.}
    \label{fig:deuteron_flux_WW}
\end{figure}

The expected flux of light antinuclei at Earth from DM annihilations and
secondary production can now be estimated by employing the two-zone
propagation model and the force-field approximation, using the source
spectra computed previously as input. For concreteness, we consider only the
Einasto
DM density profile. The antideuteron flux obtained with the new coalescence
model, using the four sets of parameters for the
diffusion model, is shown in Figs.~\ref{fig:deuteron_flux_bb}
and \ref{fig:deuteron_flux_WW}. Additionally, we use the upper limit on
the annihilation cross section imposed by the AMS-02 antiproton data as
constraint. The shaded areas correspond to the expected
sensitivity for the GAPS long duration balloon flight (105 days) (yellow)
and 10-year data-taking of AMS-02
(purple)~\cite{Giovacchini:2007dwa,Aramaki:2015laa,aramaki_review_2016}.
We find that the predicted antideuteron flux can be, for optimistic parameters,
close to the sensitivity of these two experiments.


\begin{figure}[htb]
    \centering
    \includegraphics[scale=\scale]{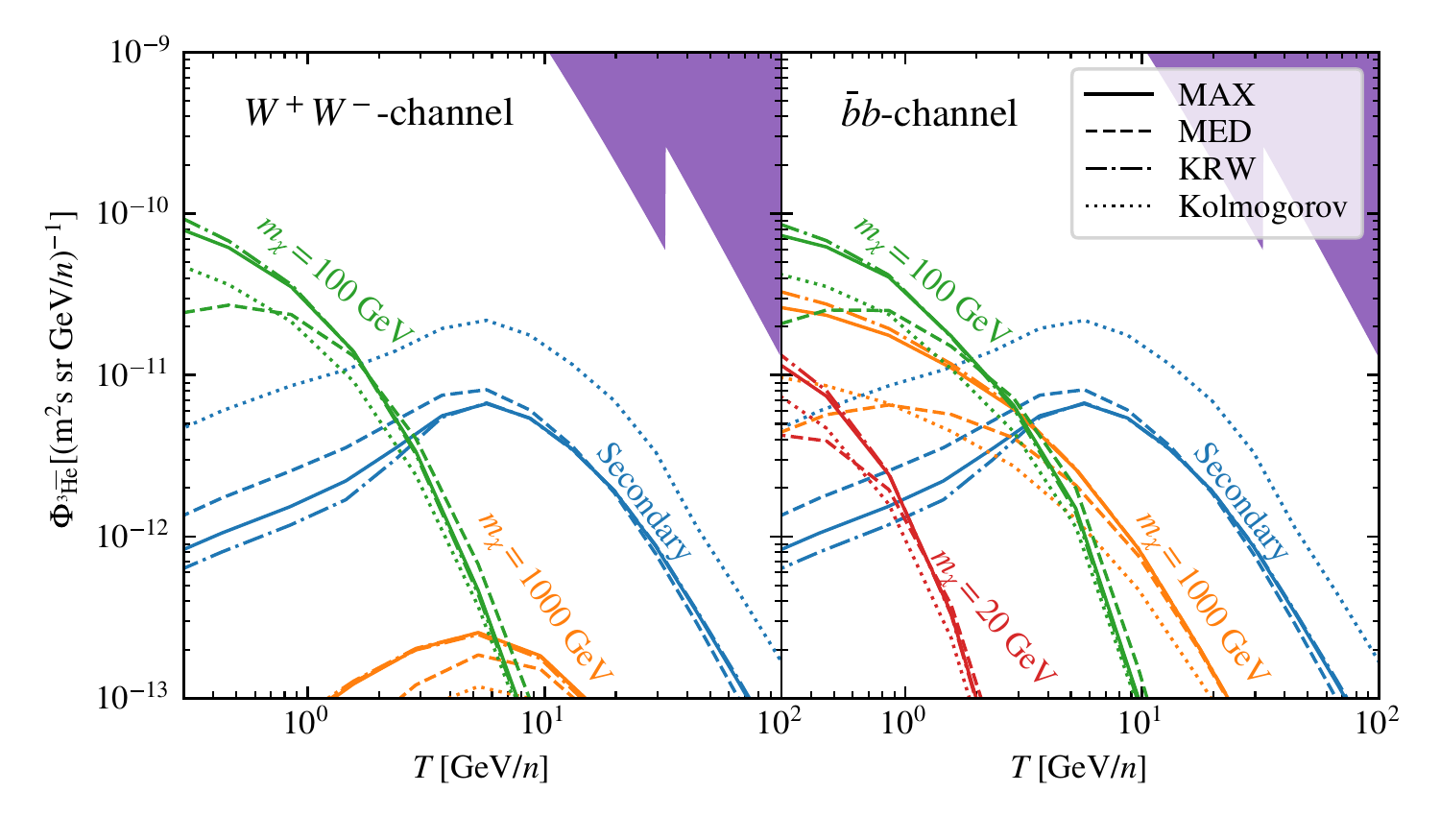}
    \caption{Estimated antihelium-3 flux on Earth from WIMP annihilations and secondary production for the considered benchmark cases and the considered propagation parameters. The shaded area on the top right is the estimated 10-year AMS-02 sensitivity.}
    \label{fig:helium_flux}
\end{figure}

The estimated antihelium-3 flux on Earth for the same benchmark cases as
in the antideuteron case is shown in Fig.~\ref{fig:helium_flux}. The
antihelium-3 sensitivity of AMS-02 is estimated by multiplying the 18-year
$\mathrm{^3\bar{He}/He}$ sensitivity from Ref.~\cite{Kounine:2010js}
with the helium flux measured by AMS-02~\cite{Aguilar:2015ctt}, and
is further rescaled to the 10-year sensitivity. The
better sensitivity for antihelium-3  than for antideuteron may explain
why AMS-02 has reported eight antihelium candidates, while the number of
antideuteron candidate events is still unknown. From
Fig.~\ref{fig:helium_flux}, one can see that antihelium nuclei from secondary
production are more likely to be detected than from DM annihilations.

There have been various other recent works investigating the detection
prospects of antihelium-3~\cite{Cirelli:2014qia,Carlson:2014ssa,Coogan:2017pwt,Blum:2017qnn,Li:2018dxj,Poulin:2018wzu,Cholis:2020twh}. The range of $p_0$
values  used in these works varies considerably, depending e.g.\ on the data
sets
used for the calibration~\cite{Li:2018dxj,Cirelli:2014qia,Carlson:2014ssa},
the hadronisation model and the event generator~\cite{Dal:2012my,Dal:2014nda}.
Since the yield of antinuclei scales as $p_0^{3A-3}$, a relatively modest
increase of $p_0$ can boost the predictions towards the experimental
sensitivities.
Alternatively, it may be a promising avenue to investigate, if modified
propagation models allow higher antideuteron and antihelium-3 fluxes, without 
being in a conflict with other observations. For instance,
Ref.~\cite{Cholis:2020twh} proposed that strong re-acceleration can increase
the number of expected antideuteron and antihelium-3 events considerably.

\section{Summary and conclusions}
\label{sec:summary}

The coalescence momentum $p_0$ of the usually employed coalescence models
in momentum space is a free parameter that must be fitted to experimental data.
Although $p_0$ should be independent on both the center-of-mass energy of the 
collision and the reaction type, the value obtained by fitting the model to
data from different reactions varies considerably~\cite{aramaki_review_2016,ibarra_prospects_2013,dal_alternative_2015,Gomez-Coral:2018yuk}.
In contrast, we have shown that the single parameter $\sigma$ of our
alternative coalescence model is universal, agreeing numerically well with its
interpretation as the size of the formation region of antinuclei.
Therefore, the production of antideuteron and antihelium-3 can be
described successfully both for point-like interactions ($e^+e^-$, DM decays and
annihilations) and for hadronic and nuclear interactions, using a single
free parameter.

Combining our coalescence model with the event generator QGSJET-II-04m,
we have calculated consistently the yield of antideuterons in
proton-proton, proton-helium, helium-helium, antiproton-proton and
antiproton-helium collisions. Thereby we avoided the use of a nuclear
enhancement factor, which is generally
ill-defined~\cite{Kachelriess:2014mga,Kachelriess:2015wpa}.
In particular, we found that the low energy tail of the secondary source
spectrum of antinuclei is strongly dominated by the contributions of
proton-helium
and helium-helium collisions. This is in contrast to previous works using
a nuclear enhancement factor, which found that antiproton collisions
should be dominant due to the low threshold energy. 
Moreover, our new coalescence model
takes into account the increase of the size of the formation region of
antinuclei in reactions involving helium,
an effect which is neglected using the old coalescence model. 
Therefore the old treatment tends to over-predict
the antideuteron yield in reactions involving helium.

Using a two-zone diffusion model, we derived the resulting fluxes of
antideuterons and antihelium. Our results indicate that  no antihelium
nuclei from secondary production or from WIMP annihilations should be
detected during 10-years of operation of AMS-02 and the long duration
balloon flights of GAPS. In contrast, the antideuteron flux can be close
to the sensitivities of the AMS-02 and GAPS experiments. Since our analysis
contains several sources of uncertainties related to,  e.g.\ the propagation
model, nuclear cross sections, and the coalescence model, the true fluxes
might be well higher and thus in reach of these experiments. We note also
that updated sensitivity analyses for both antideuteron and antihelium are
highly warranted. The GAPS experiment is most sensitive to low-energy
antideuterons from light DM. In that energy range, a more complete
numerical treatment of the CR propagation would be desirable. 
In the case of antihelium-3, the contribution from CR interaction on gas
is closer to the expected sensitivity than from DM annihilations. An
interesting avenue to investigate is whether modified
propagation models allow higher antihelium-3 fluxes without 
being in a conflict with other observations.

\section*{Acknowledgements}
\noindent
S.O.\ acknowledges support from the project OS\,481/2-1 of the 
Deutsche Forschungsgemeinschaft.


\appendix

\section{Experimental data used}
\label{sec:experiments}

We only consider experiments on antideuteron production, i.e., we neglect
the experimental data on deuteron production. In this way, we avoid 
possible contaminations from the production of deuterons in the detector
material.

\subsection{$e^+e^-$ annihilations}

The ALEPH~\cite{collaboration_deuteron_2006} and OPAL collaboration~\cite{opal_collaboration_search_1995} 
at LEP measured the deuteron and the antideuteron fluxes
 in $e^+e^-$ collisions at the $Z$ resonance. The ALEPH collaboration measured a production
 of $(5.9\pm1.8\pm 0.5)\times 10^{-6}$ antideuterons per hadronic $Z$ decay in the 
antideuteron momentum range $0.62<p<1.03\unit{GeV}$ and a production angle $|\cos\theta|<0.95$. 
Here, the first uncertainty is the statistical and the second one is the systematic error. In contrast, the OPAL collaboration did not detect any antideuterons in the momentum range $0.35<p<1.1\unit{GeV}$. We take the resulting upper limit into account by following the procedure discussed in Ref. \cite{dal_alternative_2015}.

\subsection{Proton-proton collisions}

The ALICE collaboration measured the invariant differential yield $E\dv*[3]{n}{p}$ of 
antideuterons, antitritium and antihelium-3 in inelastic proton-proton collisions 
at centre of mass energies $\sqrt{s}=\{0.9, 2.76, 7\}\unit{TeV}$ in the $p_T$ range 
$0.8\unit{GeV}<p_T<3\unit{GeV}$ and rapidity range $|y|<0.5$ \cite{collaboration_production_2018}. 
The experiment included a trigger (V0) that required a hit (charged particle) in both 
of the two pseudo-rapidity ranges $2.8<\eta <5.1$ and $-3.7<\eta <-1.7$, used to select 
non-diffractive inelastic events. We generate inelastic events and only include those
which satisfy the V0 trigger. 

The inclusive differential cross section of antideuterons at $\theta_\mathrm{cm}=90^\circ$ 
($y=0$) in $\sqrt{s}=53\unit{GeV}$ $pp$ collisions was measured at CERN ISR \cite{Alper:1973my,Henning:1977mt}. 
We compute the differential cross section as 
$E\dv*[3]{\sigma}{p}=\sigma_\mathrm{inel}/(2\pi p_TN_\mathrm{inel})(\dv*{^2N}{p_T\dd{y}})$ 
and require that $|y|<0.1$.

\subsection{Proton-beryllium and proton-aluminium  collisions}

The production of $d$, $t$, $^3\mathrm{He}$, $\bar d$, $\bar t$ and
$\bar{^3\mathrm{He}}$ at $0^\circ$ with momenta between 12 and 37$\unit{GeV}$ in
the lab frame in $p$-beryllium and $p$-aluminium collisions at
$200\unit{GeV}/c$ was reported in Ref. \cite{Bozzoli:1979fh}. The results are presented as ratios of antinuclei and $\pi^-$ yields. The antideuteron results had been split into three and five bins between 20 and $37\unit{GeV}$ in $p$-aluminium and $p$-beryllium collisions, respectively. As the data are given for $0^\circ$ in the lab frame, and we are only interested in the bulk of produced antinuclei, we include all produced $\pi^-$ and antideuterons in the analysis.

\section{Parametrisation of the primary cosmic ray flux}
\label{app:parametrisation}

In order to describe the secondary production of antinuclei, one needs 
the primary fluxes of protons, helium and antiprotons as input. The primary CR
fluxes were traditionally parametrised by an unbroken power law up to the
CR knee, as $\Phi(T)\propto T^{-\gamma}$, where $T$ is the kinetic energy of
the particle and $\gamma\sim 2.7$. However, recent experimental data, such as
from the AMS-02~\cite{aguilar_antiproton_2016,Aguilar:2015ctt,Aguilar:2015ooa} and CREAM~\cite{Yoon:2017qjx} experiments, clearly suggest that there is a hardening in the CR flux around the rigidity $\mathcal{R}\sim 400 \unit{GeV}$. In
addition, there are now several experiments, including CREAM and
DAMPE~\cite{An:2019wcw}, suggesting that there is an additional break around
10\,TeV. For a spectrum with $N$ statistical significant breaks, we fit the
function
\begin{equation}
    \Phi(T)=AT^{-\gamma}\left(\frac{T}{T + b}\right)^c\prod_{i=1}^Nf(T_{\mathrm{b}i}, \Delta\gamma_i, s),
    \label{eq:parametrisation}
\end{equation}
where
\begin{equation}
f(T_{\mathrm{b}}, \Delta\gamma, s)= \left[1 + \left(\frac{T}{T_\mathrm{b}}\right)^{s}\right]^{\Delta\gamma/s}    
\end{equation}
accounts for the breaks, while the first parentheses is included to reproduce
the low energy part of the spectra. We follow Ref.~\cite{An:2019wcw} and fix
the smoothness parameter $s=5$ for proton and antiproton, while we find that
$s=3$ provides a good fit for helium. The parameters $\Delta\gamma_i$ describe
the changes in the power-law index. Thus, for each additional break, we add
two free parameters, while for the main spectrum, we have four free parameters. We fix the parameters by first fitting
\begin{equation}
    \Phi_\text{AMS-02}(T)=AT^{-\gamma}\left(\frac{T}{T + b}\right)^c,
\label{eq:ams_param}
\end{equation}
to the AMS-02 proton data up to the hardening at $T\sim400\unit{GeV}/n$ and
in turn fix the remaining parameters by using Eq.~\eqref{eq:parametrisation}.

We take into account solar modulation by using the force field
approximation~\cite{gleeson_solar_1968}. Based on the Oulu NM
database~\cite{database_phi}, we find the mean solar modulation force-field
$\phi$ in the periods of data taking~\cite{2015ApJ...815..119V,Usoskin:2017cli}.
Since solar modulation can be neglected at high energies, this is only
relevant for the AMS-02 data. For the proton and helium fluxes, we obtained
$\phi=0.60$\,GV, while we found $\phi=0.62$\,GV for the antiproton flux. Our
fit results are listed in Table~\ref{tab:params}.

\begin{table}
\centering
    \begin{tabular}{lcccccccccc}
        \toprule
         & $A$ & $b$ & $c$ & $\gamma$ & $T_\mathrm{b1}$ & $\Delta\gamma_1$ & $T_\mathrm{b2}$ & $\Delta\gamma_2$ &  $\chi^2/\mathrm{d.f.}$ \\
         &\hspace{-1.2cm}[$\mathrm{m^2s\;sr\;/(GeV}/n)$] & \hspace{-.3cm}[$\mathrm{GeV}/n$] &  & & \hspace{-.3cm}[$\mathrm{GeV}/n$] & & \hspace{-.3cm}[$\mathrm{GeV}/n$] & & \\
        \midrule
        Proton & 26714 & 0.49 & 6.81 & 2.88 & 343 & 0.265 & 19503 & -0.264 & 0.39 \\
        Helium & 1151 & 1.06 & 2.74 & 2.79 & 237 & 0.309 & 18849 & -0.620 & 0.95 \\
        Antiproton & 22.4 & 1.28 & 9.22 & 3.22 & 88.4 & 0.412 & & & 0.39 \\
        \bottomrule
    \end{tabular}
\caption{Parameters for the fits of the primary proton, helium, and antiproton spectra at local interstellar space.
 The fitting procedure is discussed in the text.}
\label{tab:params}
\end{table}


\providecommand{\href}[2]{#2}\begingroup\raggedright\endgroup

\end{document}